\documentclass[iop,apj,tighten]{emulateapj}
\usepackage{natbib}
\usepackage{amsmath,amstext}
\usepackage{graphicx} 
\usepackage{multirow}
\usepackage{comment}

\usepackage{amssymb}

\newcommand{\degrees}{$\!\!$\char23$\!$}

\begin{document}

\title[NGC 5053: Orbit, Mg, Al, and Si]{The Metal-poor non-Sagittarius (?) Globular Cluster NGC 5053: \\Orbit and Mg, Al and Si Abundances}
\author{Baitian Tang\altaffilmark{1,2}}
\author{J. G. Fern\'andez-Trincado\altaffilmark{2,10}}
\author{Doug Geisler\altaffilmark{2}}
\author{Olga Zamora\altaffilmark{3,4}}
\author{Szabolcs M\'esz\'aros\altaffilmark{5,17}}
\author{Thomas Masseron\altaffilmark{3,4}}
\author{Roger E. Cohen\altaffilmark{6}}
\author{D. A. Garc\'ia-Hern\'andez\altaffilmark{3,4}}
\author{Flavia Dell'Agli\altaffilmark{3,4}}
\author{Timothy C. Beers\altaffilmark{7}}
\author{Ricardo P. Schiavon\altaffilmark{8}}
\author{Sangmo Tony Sohn\altaffilmark{6}}
\author{Sten Hasselquist\altaffilmark{9}}
\author{ Annie C. Robin\altaffilmark{10}}
\author{Matthew Shetrone\altaffilmark{11}}
\author{Steven R. Majewski\altaffilmark{12}}
\author{Sandro Villanova\altaffilmark{2}}
\author{Jose Schiappacasse Ulloa\altaffilmark{2}}
\author{Richard R. Lane\altaffilmark{13}}
\author{Dante Minnti\altaffilmark{13}}
\author{Alexandre Roman-Lopes\altaffilmark{15}}
\author{Andr\'es Almeida\altaffilmark{15}}
\author{E. Moreno\altaffilmark{16}}

\affil{$^1$School of Physics and Astronomy, Sun Yat-sen University, Zhuhai 519082, China}
\affil{$^2$Departamento de Astronom\'{i}a, Casilla 160-C, Universidad de Concepci\'{o}n, Concepci\'{o}n, Chile}
\affil{$^3$Instituto de Astrof\'isica de Canarias, 38205 La Laguna, Tenerife, Spain}
\affil{$^4$Departamento de Astrof\'isica, Universidad de La Laguna, 38206 La Laguna, Tenerife, Spain}
\affil{$^5$ELTE E\"otv\"os Lor\'and University, Gothard Astrophysical Observatory, Szombathely, Hungary}
\affil{$^6$Space Telescope Science Institute, 3700 San Martin Drive,
Baltimore, MD 21218, USA}
\affil{$^7$Department of Physics and JINA Center for the Evolution of the Elements,
University of Notre Dame, Notre Dame, IN 46556  USA}
\affil{$^8$Astrophysics Research Institute, Liverpool John Moores University, Liverpool, United Kingdom}
\affil{$^9$New Mexico State University, Las Cruces, NM 88003, USA}
\affil{$^{10}$Institut Utinam, CNRS UMR 6213, Universit\'e Bourgogne-Franche-Comt\'e, OSU THETA Franche-Comt\'e, Observatoire de Besan\c{c}on, BP 1615, 25010 Besan\c{c}on Cedex, France}
\affil{$^{11}$Department of Astronomy, University of Texas at Austin, Austin, TX 78712, USA}
\affil{$^{12}$Department of Astronomy, University of Virginia, Charlottesville, VA 22904-4325, USA}
\affil{$^{13}$Instituto de Astrofisica, Pontificia Universidad Cat\'olica de Chile, Av. Vicuna Mackenna 4860, 782-0436 Macul, Santiago, Chile}
\affil{$^{14}$Department of Physics and Astronomy, Universidad de La Serena, Cisternas 1200, La Serena, Chile}
\affil{$^{15}$Instituto de Investigaci\'on Multidisciplinario en Ciencia y Tecnolog\'i­a, Universidad de La Serena, Benavente 980, La Serena, Chile} 
\affil{$^{16}$Instituto de Astronom\'ia, Universidad Nacional Aut\'onoma de M\'exico, Apartado Postal 70-264, 04510-M\'exico DF, Mexico}
\affil{$^{17}$Premium Postdoctoral Fellow of the Hungarian Academy of Sciences}





\begin{abstract}
Metal-poor globular clusters (GCs) exhibit intriguing Al-Mg anti-correlations and possible Si-Al correlations, which are important clues to decipher the multiple-population phenomenon. NGC 5053 is one of the most metal-poor GCs in the nearby Universe, and has been suggested to be associated with the Sagittarius (Sgr) dwarf galaxy, due to its similarity in location and radial velocity with one of the Sgr arms. In this work, we simulate the orbit of NGC 5053, and argue against a physical connection between Sgr and NGC 5053. On the other hand, the Mg, Al, and Si spectral lines, which are difficult to detect in the optical spectra of NGC 5053 stars, have been detected in the near-infrared APOGEE spectra. We use three different sets of stellar parameters and codes to derive the Mg, Al, and Si abundances. Regardless of which method is adopted, we see a large Al variation, and a substantial Si spread. Along with NGC 5053, metal-poor GCs exhibit different Mg, Al, and Si variations. Moreover, NGC 5053 has the lowest cluster mass among the GCs that have been identified to exhibit an observable Si spread until now.

\end{abstract}

\maketitle

\section{Introduction}
\label{sect:intro}

The multiple-population (MP) phenomenon has now been observed in most of the evolutionary sequences of globular clusters (GCs): main-sequence (MS), subgiant branch (SGB), red giant branch (RGB), horizontal branch (HB), and asymptotic giant branch (AGB). While photometry is an efficient way of revealing MPs in most stellar phases, and particularly, fainter phases \citep[e.g.,][]{Milone2015, Piotto2015}, high-resolution spectroscopy allows deeper insights into GC formation and internal stellar evolution, providing detailed abundances for a number of elements with a variety of nucleosynthetic origins \citep[e.g.,][]{GH2015,Meszaros2015, Schiavon2017, Tang2017}.  The Al$-$Mg anti-correlations, and the less frequent Si$-$Al correlations are of great interest, because these chemical patterns indicate that the polluting stars must have reached temperatures above 90 MK \citep{Prantzos2007, Ventura2012, DAntona2016}, which is a strong constraint on the nature of these stars. A significant silicon spread is observed in a few GCs, mainly metal-poor or massive GCs \citep[e.g.,][]{Carretta2009b, Meszaros2015}. Further investigations of metal-poor GCs would tell us more about the environment that stimulates the Si variation.

\begin{figure}
\centering
\includegraphics [width=0.45\textwidth]{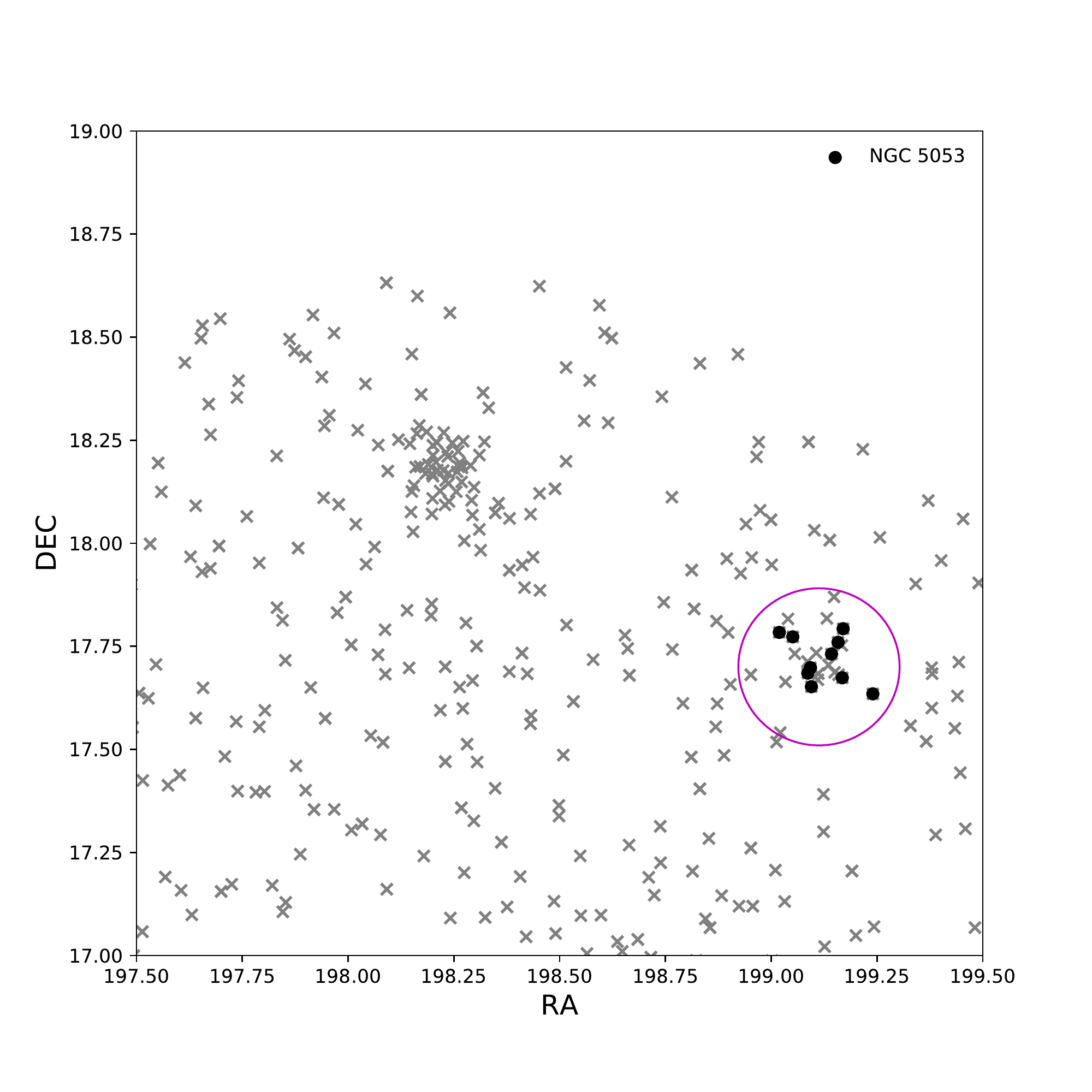} 
\caption{Locations of the stars in the APOGEE field of NGC 5053. The NGC 5053 member stars are indicated by filled circles. The magenta circle indicates the cluster tidal radius. Note that the star cluster in the upper left is M53.}\label{fig:loc}
\end{figure}

NGC 5053 is one of the most metal-poor GCs known ([Fe/H$]=-2.27$, \citealt{Harris1996}; 2010 version). It is located close to the Galactic north pole ($\alpha_{\rm J2000} =  \rm 13^h16^m27.1^s$, $\delta_{\rm J2000} = +17^{\circ}42'01''$, $l=335.70^{\circ}$, $b = +78.95^{\circ}$). With similar location and radial velocity\footnote{The radial velocities in this work are heliocentric radial velocities.} (RV) to one of the Sagittarius (Sgr) arms \citep{Law2010}, NGC 5053 has been speculated to be associated with the Sgr stream, but a firm association is still under debate \citep{Palma2002, Bellazzini2003,Law2010}. A similar discussion is also found for its close companion, M53 (Figure \ref{fig:loc}).
NGC 5053 and M53 have similar Galactocentric radius (17.8 and 18.4 kpc, respectively). Based on the X, Y, Z positions from the Harris catalog, these two GCs are only $\sim$500 pc away from each other currently. Furthermore, \citet{Chun2010} suggested that these two GCs also share a tidal bridge.
 \citet{Palma2002} suggested the Sgr association is unlikely for these two GCs because they are more metal poor than previously identified clusters ([Fe/H$]=-2.27$ for NGC 5053 and [Fe/H$]=-2.10$ for M53, Harris catalog). \citet{Bellazzini2003} considered NGC 5053 as one of the Sgr GCs based on its location and radial velocity, while M53 was not discussed. \citet{Law2010} showed that the proper motion of M53 along the right ascension axis ($\mu_{\alpha}cos\delta$) is not consistent with any of the Sgr tidal arms, and concluded that M53 is probably not physically associated with Sgr. Meanwhile, the location and radial velocity of NGC 5053 are consistent with the T1 Sgr tidal arm, but the cluster's proper motion was not available at that time. 
Below, we simulate the orbit of NGC 5053 using newly obtained proper motion\footnote{The proper motions in this work are absolute proper motions.} and our new software for dynamical modeling GravPot16 (Fern\'andez-Trincado et al., in prep.).

In terms of chemical abundances, \citet{Sbordone2015} analyzed HIRES spectra ($438-678$ nm, $R=48\, 000$) downloaded from the Keck Observatory archive for one RGB star in NGC 5053. Because the spectral lines are generally weak in metal-poor stars, only 13 element species were investigated. We note that Mg, Al, and Si are not in their list of elements. \citet[B15]{Boberg2015} observed 17 member stars using the Wisconsin-Indiana-Yale-NOAO-Hydra (WIYN) multi-object spectrograph. After analyzing the medium-resolution spectra ($R\sim 13\,000$), they measured the abundances of Fe, Ca, Ti, Ni, Ba, Na, and O. Boberg et al. found bimodality in the Na distribution, suggesting the existence of multiple populations in this GC. Previous work on elemental abundances provides evidence for a possible chemical connection between NGC 5053 and Sgr, including a very low yttrium abundance \citep{Sbordone2015}. However, a firm association has not been obtained yet. In this work, we focus on deriving the Mg, Al, and Si abundances from the APOGEE (Apache Point Observatory Galactic Evolution Experiment) spectra (Section \ref{sect:chem}), where relatively strong spectral lines are seen for these elements. These are the first measurements of Mg, Al, and Si in NGC 5053. These abundances are important indicators of the nature of the polluting stars that are responsible for multiple populations (Section \ref{sect:dis}).

\begin{figure}
\centering
\includegraphics [width=0.45\textwidth]{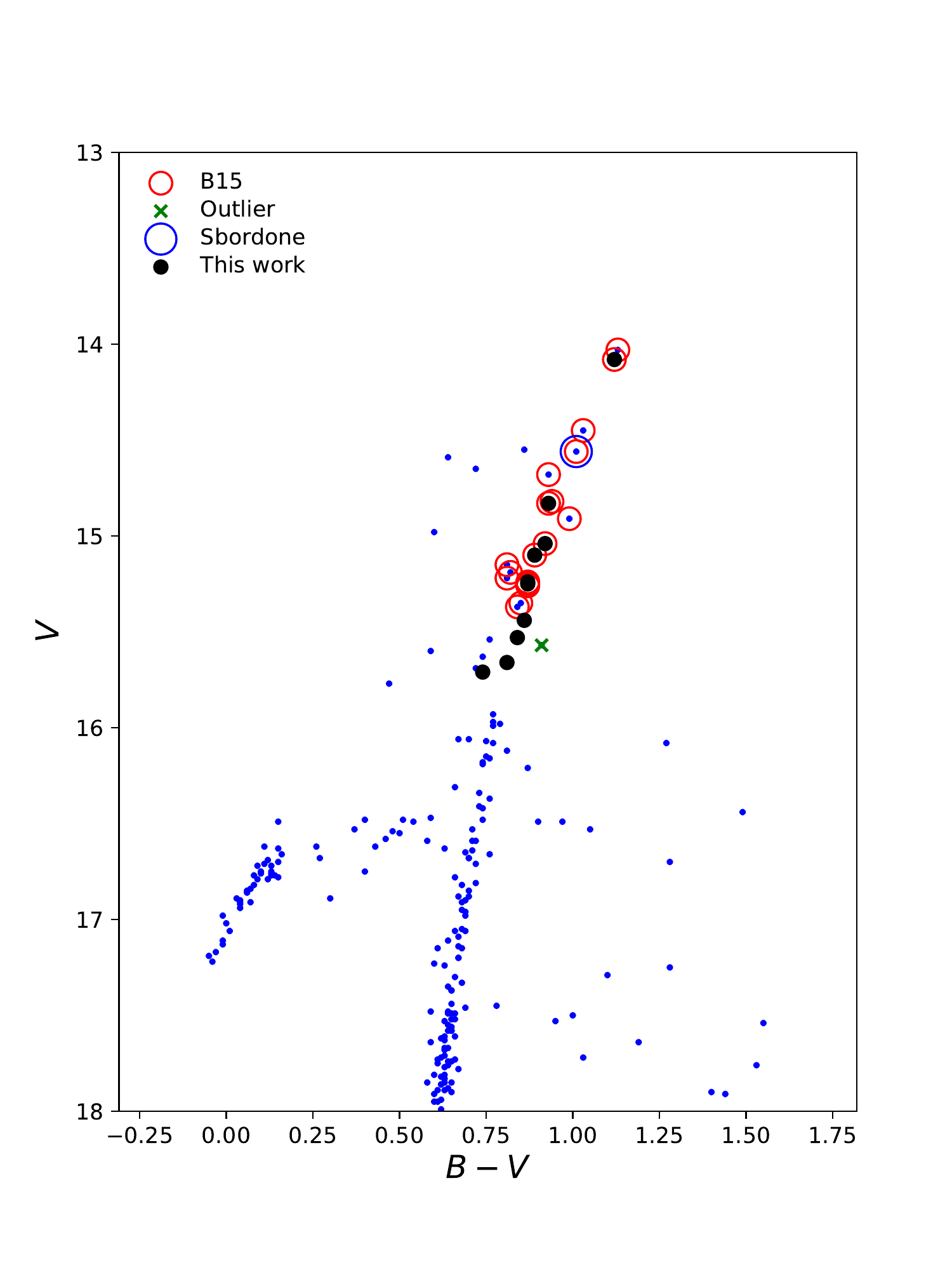} 
\caption{V versus B-V CMD of NGC 5053. The APOGEE selected members are indicated by black dots. The red open circles show the stars observed by B15. The blue open circle is the one RGB star observed in \citet{Sbordone2015}. The cross is the one star that we exclude by CMD and isochrone selection.}\label{fig:cmd}
\end{figure}

\begin{table*}
\caption{Photometric Properties of the NGC 5053 Members.}              
\label{tab:pho}      
\centering                                      
\begin{tabular}{c c c c c c c c c c}         
\hline\hline     
\# & APOGEE\_ID & RA$_{\rm J2000}$ (\degrees) & DEC$_{\rm J2000}$ (\degrees) & $B$ &  $V$ &  2MASS $J$ & 2MASS $H$ & 2MASS $K_S$ & B15 ID\\
\hline
Star1  & 2M13160457+1747017 & 199.019064 & 17.783821 & 16.37 & 15.53 & 13.775 & 13.208 & 13.116 & ...\\
Star2  & 2M13161223+1746228 & 199.050959 & 17.773018 & 15.20 & 14.08 & 11.956 & 11.377 & 11.243 & 3\\
Star3  & 2M13162073+1741059 & 199.086404 & 17.684973 & 15.99 & 15.10 & 13.293 & 12.752 & 12.633 & 15\\
Star4  & 2M13162226+1741536 & 199.092771 & 17.698231 & 16.12 & 15.25 & 13.140 & 12.776 & 12.608 & 20\\
Star5  & 2M13162279+1739074 & 199.094981 & 17.652067 & 16.30 & 15.44 & 14.101 & 13.655 & 13.513 & ...\\
Star6  & 2M13163419+1743530 & 199.142482 & 17.731409 & 16.45 & 15.71 & 13.471 & 12.873 & 12.856 & 19\\
Star7  & 2M13163790+1745355 & 199.157922 & 17.759867 & 16.11 & 15.24 & 13.180 & 12.637 & 12.541 & 14\\
Star8  & 2M13164030+1740254 & 199.167950 & 17.673727 & 15.96 & 15.04 & 13.926 & 13.424 & 13.348 & ...\\
Star9  & 2M13164077+1747338 & 199.169899 & 17.792746 & 16.47 & 15.66 & 13.345 & 12.793 & 12.688 & ...\\
Star10  & 2M13165764+1738050 & 199.240180 & 17.634743 & 15.76 & 14.83 & 12.962 & 12.388 & 12.295 & 11\\
\hline                                             
\end{tabular}

\raggedright{Note: $B$ and $V$ apparent magnitudes come from \citet{Sarajedini1995}. }\\
\end{table*}

\section{Sample Selection and Data Reduction}
\label{sect:data}

APOGEE \citep{Majewski2017} was one of the programs operating during the Sloan Digital Sky Survey III (SDSS-III, \citealt{Eisenstein2011}). The multi-object NIR fiber spectrograph on the 2.5 m telescope at Apache Point Observatory \citep{Gunn2006} delivers high-resolution ($R\sim$22,500) $H$-band spectra ($\lambda = 1.51 - 1.69$ $\mu$m), and the APOGEE survey targeted a color-selected sample that predominantly consists of RGB stars across the Milky Way \citep{Zasowski2017}.  APOGEE data reduction software is applied to reduce multiple 3D raw data cubes into calibrated, well-sampled, combined 1D spectra \citep{Nidever2015}. In addition, the APOGEE Stellar Parameter and Chemical Abundances Pipeline (ASPCAP; \citealt{GP2016}) derives stellar parameters and elemental abundances by comparing observed spectra to libraries of theoretical spectra \citep{Zamora2015}, constructed using extensive molecular/atomic linelists \citep{Shetrone2015}, in order to find the closest model match, using $\chi^2$ minimization in a multidimensional parameter space. In the current public SDSS Data Release 14 (DR14), up to more than 20 chemical elements were identified, and measured abundances provided (\citealt{Holtzman2015}; in prep.). 
DR14 includes all APOGEE-1 data, and APOGEE-2N data taken between July 2014 and July 2016. 

To select stars that are members of NGC 5053, we first restrict the sample to stars within the tidal radius (11.43') given by Harris (1996; 2010 version). Then we exclude probable field stars by applying two generous criteria: [Fe/H$]_{\rm NGC5053}\pm0.3$, RV$_{\rm NGC5053}\pm15$ km/s, where  [Fe/H$]_{\rm NGC5053}=-2.27$ and RV$_{\rm NGC5053}=44.0$ km/s from the Harris catalog. We identify 11 stars using the metallicities and RVs provided by APOGEE DR14. Given that NGC 5053 may have tidal structure \citep{Lauchner2006}, we also relax the search area to 3 times the tidal radius, but no further members in DR14 are identified in this way. Since NGC 5053 is far away from the Galactic plane, relaxing the [Fe/H] and RV criteria ($\pm0.6$ dex and $\pm 30$ km/s) also do not increase the number of members. After checking the color-magnitude diagram (CMD) and stellar parameters, we further exclude one star (APOGEE\_ID=2M13161337+1743552), because its location in the CMD is away from the RGB (Figure \ref{fig:cmd}), and its ASPCAP stellar parameters also deviate from the isochrone (Figure \ref{fig:iso}). The stellar parameters are given in Section \ref{sect:dev}. Thus, our final list consists of 10 stars (Figure \ref{fig:loc}). We have also checked the photometry from P. Stetson\footnote{http://www.cadc-ccda.hia-iha.nrc-cnrc.gc.ca/en/community/STETSON/standards/} for possible AGB (Asymtoptic Giant Branch) stars in our sample. \cite{GH2015} showed that $U$ vs. $(U-I)$ CMD is able to distinguish AGB stars from RGB stars. We have checked the $U$ vs. $(U-I)$ CMD for NGC 5053, where eight stars of our sample are identified. No obvious AGB star is found.  Among our ten sample stars, six of them were also selected as members in B15 (Table \ref{tab:pho}, also see Figure \ref{fig:cmd}).

\section{TRACKING THE ORBIT OF NGC 5053}
\label{sect:orb}

%

\begin{figure*}
\centering
\includegraphics [angle=0,width=0.45\textwidth]{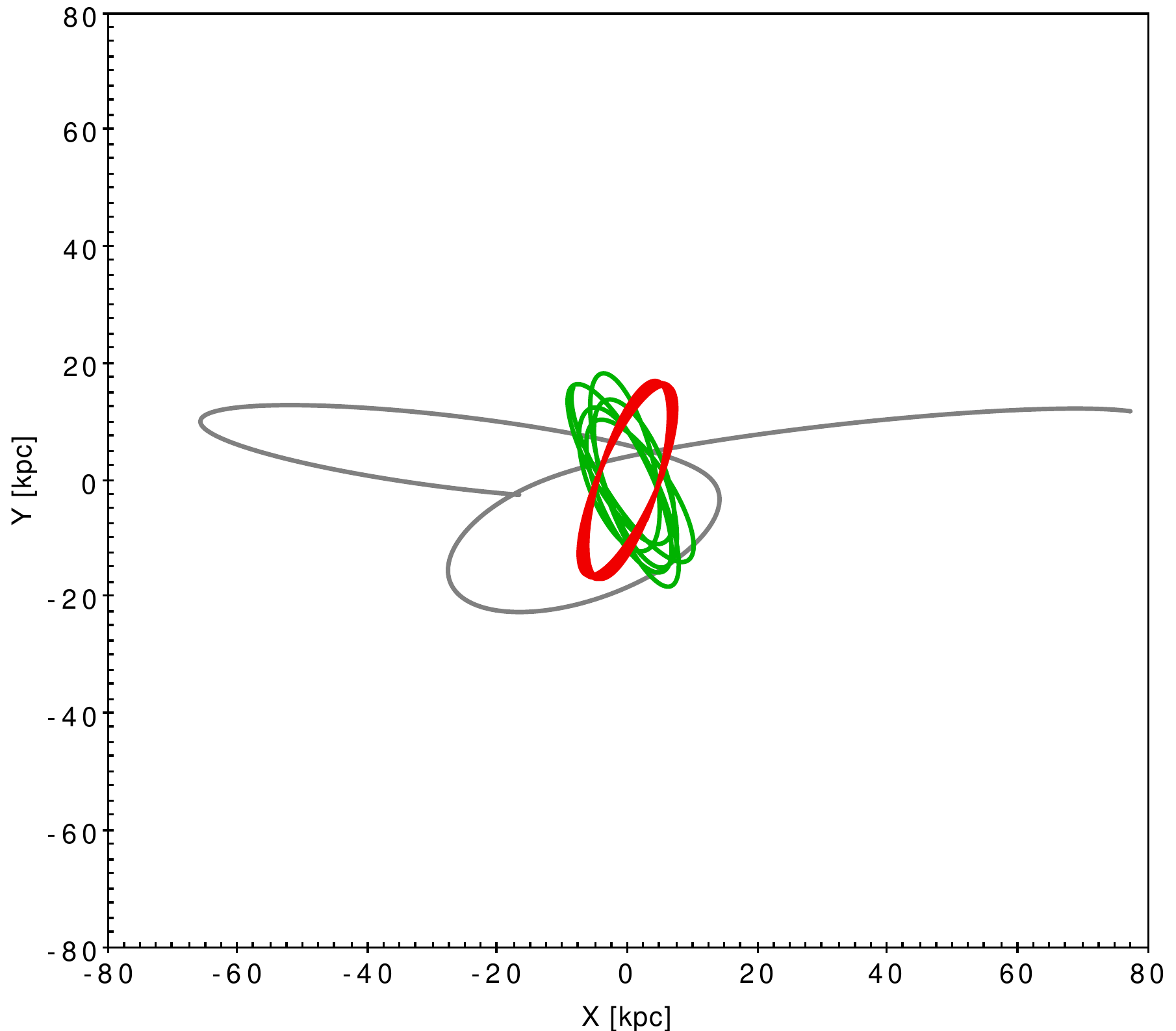} 
\includegraphics [angle=0,width=0.45\textwidth]{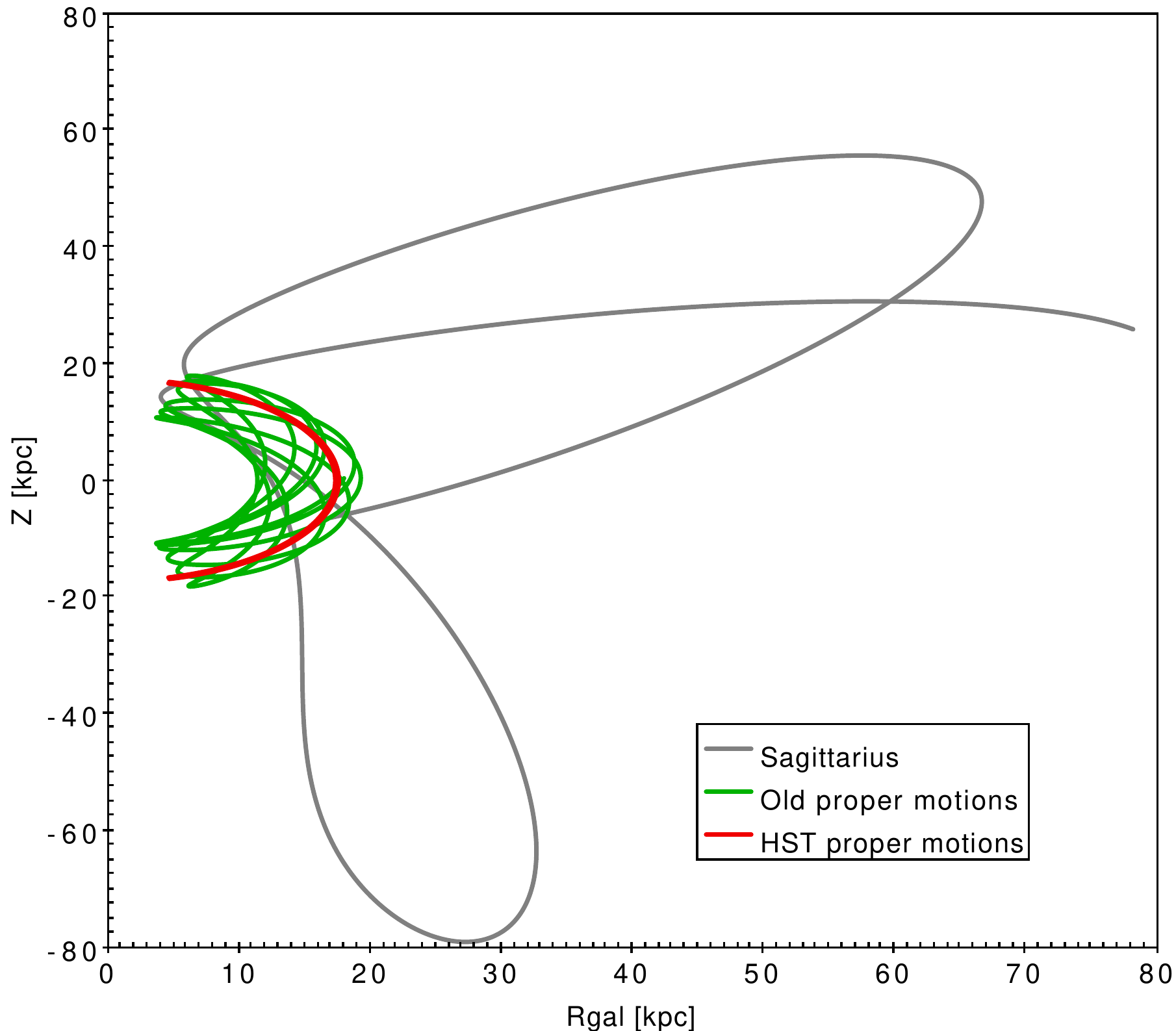} 
\caption{ Representative orbits of Sgr (grey), NGC 5053 with HST proper motion (red),  and NGC 5053 with old proper motion (green) in the reference frame where the bar is at rest. Left: Equatorial Galactic orbits. Right: Meridional Galactic orbits. The orbits of NGC 5053 do not resemble that of Sgr, suggesting the proposed connection between these two systems may be unlikely.}\label{fig:sim}
\end{figure*}

\begin{table*}
\caption{RVs, Distances, and Proper Motions Adopted in This Work.}              
\label{tab:input}      
\centering                                      
\begin{tabular}{c c c c c c c c c}         
\hline\hline     
\# & RV & RV Err & d & d Err &  $\mu_{\alpha}cos\delta$ &  $\mu_{\alpha}cos\delta$ Err & $\mu_{\delta}$ &  $\mu_{\delta}$ Err\\
\hline
&\multicolumn{2}{c}{(km/s)}& \multicolumn{2}{c}{(pc)} &\multicolumn{2}{c}{(mas/yr)} &\multicolumn{2}{c}{(mas/yr)}\\
\hline
Sgr&171.0 &17.0 &24.8 &0.8 &-2.75 &0.2  & -1.65 &0.22\\
NGC 5053 old proper motion&\multirow{2}{*}{43.4}  &\multirow{2}{*}{2.2}  &\multirow{2}{*}{16.7} &\multirow{2}{*}{0.3} & -5.81 &0.53 & -2.76 &0.53\\
NGC 5053 HST proper motion& & & & &\multicolumn{4}{c}{Sohn et al. (in prep.)}\\
\hline                                             
\end{tabular}

\raggedright{ Note: For NGC 5053, RV comes from APOGEE observation; distance comes from \citet{AF2010}; the old proper motion comes from \citet{Kharchenko2013}; the HST proper motion will be shown in Sohn et al. (in prep.)\\
For Sgr, RV comes from \citet{Kunder2009}; distance comes from \citet{Law2010}; proper motion comes from \citet{Pryor2010}.
}\\
\end{table*}

\subsection{Adopted observational data for NGC 5053 and Sagittarius}

 To obtain robust estimates of the orbit of NGC 5053, we employed recent high-quality data. First, we adopted the new proper motion measurements based on multi-epoch Hubble Space Telescope (HST) observations separated by $\sim 11$ years in time. Details of this measurement, including the proper motion result for NGC 5053, will be presented in a forthcoming paper (Sohn et al., in prep). In short, we utilized the same established method of \citet{Sohn2012, Sohn2013, Sohn2017} to achieve a 1-d proper motion uncertainty of $\sim 0.07 mas\>yr^{-1}$ for NGC 5053. The only other proper motion measurement found in the literature is provided by \citet{Kharchenko2013}, and for comparison, we integrate orbits using their measurement separately (see Figure \ref{fig:sim} for details).  
 

The heliocentric distance of NGC 5053 was presented in Harris catalog and \citet{AF2010} using RR Lyrae stars. Being standard candles, RR Lyrae stars are particularly suitable to derive cluster distances with a high degree of confidence. Harris catalog suggested a distance of 16.4 kpc for NGC 5053, which assumes $V$(HB$)=16.65$ \citep{Sarajedini1995}, [Fe/H$]=-2.29$, extinction ratio $R=3.0$, $E(B-V)=0.04$, and the relation $M_V=0.15[$Fe/H$]+0.80$. \citet{AF2010} derived a mean distance modulus of 16.12$\pm$0.4, which correspond to a distance of 16.7$\pm$0.3 kpc. They assumes [Fe/H$]=-1.97$, $R=3.2$, $E(B-V)=0.018$. Even though different parameters were assumed in two literature studies, the derived distances agree with each other. We adopt the latter distance 16.7$\pm$0.3 kpc in this work.

Using the APOGEE spectra, we have visually inspected the strong Mg, Al, and Si spectral lines (section \ref{sect:chem}), and list the derived RV in Table \ref{tab:asp}. We obtain a mean RV of 43.4$\pm$2.2 km/s, which is consistent with the value from the Harris catalog (44.0$\pm$0.4 km/s). The detailed values that we adopted in our simulations are shown in Table \ref{tab:input}.

\subsection{Orbits and Implications}

By combining our Milky Way potential model (see Appendix) with measurements of radial velocity, absolute proper motion, distance and sky position for NGC 5053 and Sagittarius, we ran $N_{\rm total} = 1\times10^{4}$ pairs of orbits NGC 5053 -- Sagittarius, taking into account the uncertainties in the input data considering 1$\sigma$ variations in a Gaussian Monte Carlo approach. For each generated set of parameters the orbit of NGC 5053 was computed backward in time, up to 2.5 Gyr. All orbital computations were made with a Runge-Kutta algorithm of seventh order (Fehlberg et al. 1968), with a variable time step determined at the initial conditions and conserved to 1 part in 10$^{-14}$.

From the integrated set of orbits, we compute (1) the orbital eccentricity, defined as $e= {(r_{apo} - r_{per})}/{(r_{apo} + r_{per})}$, with $r_{apo}$ and $r_{per}$ as successive apogalactic and perigalactic distances, (2) the maximun vertical amplitude $Z_{max}$ as well as (3) the orbital Jacobi constant per unit mass, $E_{J}$, and (4) the ``characteristic'' orbital energy, $(E_{min} + E_{max})/2$, as defined in \citet{Moreno2015}.

The major assumptions and limitations in our computations are: \textit{i}) we do not consider the effects of dynamical friction of NGC 5053 and Sagittarius, which is expected to bring both systems closer to the Galactic center; \textit{ii}) we ignore secular changes in our Milky Way potential over time, which are expected although the Milky Way has had a quiet recent accretion history; \textit{iii}) we ignore the possibility that the pair NGC 5053 -- Sagittarius were likely more massive in the past and have tidally lost some of that mass, which reduces the absolute probabilities we find. 

\begin{figure}
	\centering
	\includegraphics [width=0.50\textwidth]{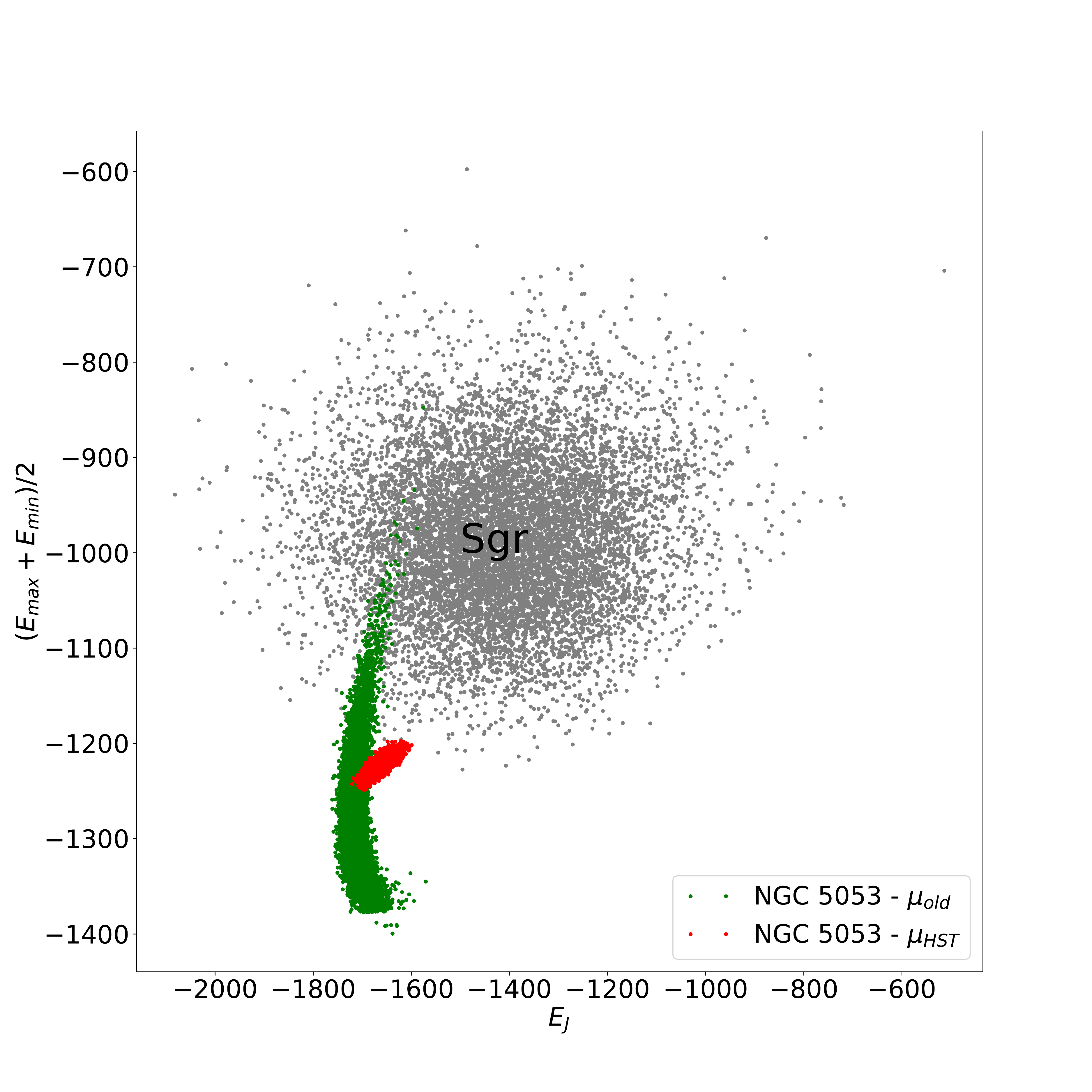} 
	\caption{This diagram plots the ``characteristic" orbital energy $(E_{min} + E_{max})/2$ versus the orbital Jacobi constant ($E_{J}$), in units of 1$\times$10$^{5}$ km$^{2}$ s$^{-2}$, in the reference frame of the bar, and considering 1$\sigma$ variations in a Gaussian Monte Carlo sampling (1$\times$10$^{4}$ orbits). Grey dots are for the Sagittarius dwarf galaxy, green dots for NGC 5053 considering the absolute proper motion from \citet{Kharchenko2013}, and red dots considering absolute proper motion from HST observations.}\label{Figure2}
\end{figure}

 Our simulation results (Figure \ref{fig:sim}) show that the orbit of NGC 5053 does not resemble that of Sgr, suggesting no association between the two objects. 
Figure \ref{Figure2} shows results for the ``characteristic'' orbital energy ($E_{max} + E_{min}/2$) versus the orbital Jacobi energy ($E_{J}$) in the potential mentioned above (including 1$\times$10$^{4}$ random orbital realisations), the Galactic non-inertial frame is employed to compute the orbits. This plot shows a noticeable difference between NGC 5053 and Sgr. The Sgr occupies a very distinct region in these spaces, with an over-density at $E_{J} \sim$-1400$\times$10$^{5}$ km$^{2}$ s$^{-2}$ and $E_{max} + E_{min}/2 \sim $ -1000  km$^{2}$ s$^{-2}$, with most of the NGC 5053 orbits shifted toward a lower distribution both in ``characteristic" orbital energy and $E_{J}$. Therefore an association of NGC 5053 with Sgr appears implausible. 

\begin{figure}
	\centering
	\includegraphics [width=0.50\textwidth]{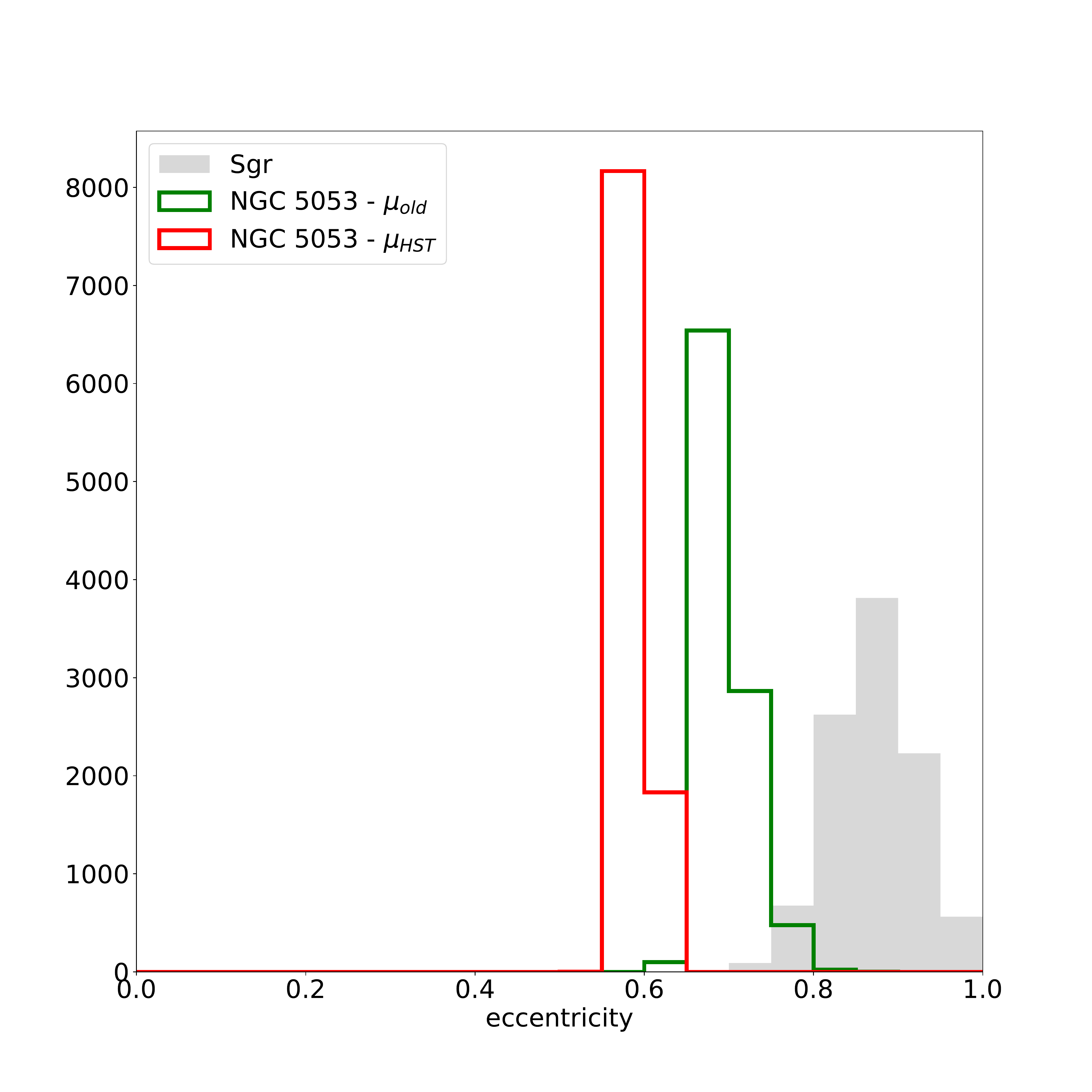} 
	\caption{Histogram of the distribution of eccentricities for NGC 5053 and Sagittarius considering 1$\sigma$ variations in a Gaussian Monte Carlo sampling (1$\times$10$^{4}$ orbits). The Sagittarius eccentricity is shown with grey histogram, while NGC 5053 adopting HST absolute proper motion are represented with a red histogram, and from adopted absolute proper motion from \citet{Kharchenko2013} is represented with a green histogram.}\label{Figure3}
\end{figure}

In Figure \ref{Figure3} we compared the eccentricity distribution for Sgr and NGC 5053 considering the HST and \citet{Kharchenko2013} absolute proper motions. This figure reassuringly shows that the eccentricity distributions of NGC 5053 and Sgr are decidedly different, even after taking errors into account, which means that NGC 5053 is characterized by different orbital configurations than Sgr. Therefore, our results suggest that it is unlikely that NGC 5053 comes from a dwarf galaxy (Sgr) that, on average, has significantly higher orbital energy.

\begin{figure}
\centering
\includegraphics [width=0.45\textwidth]{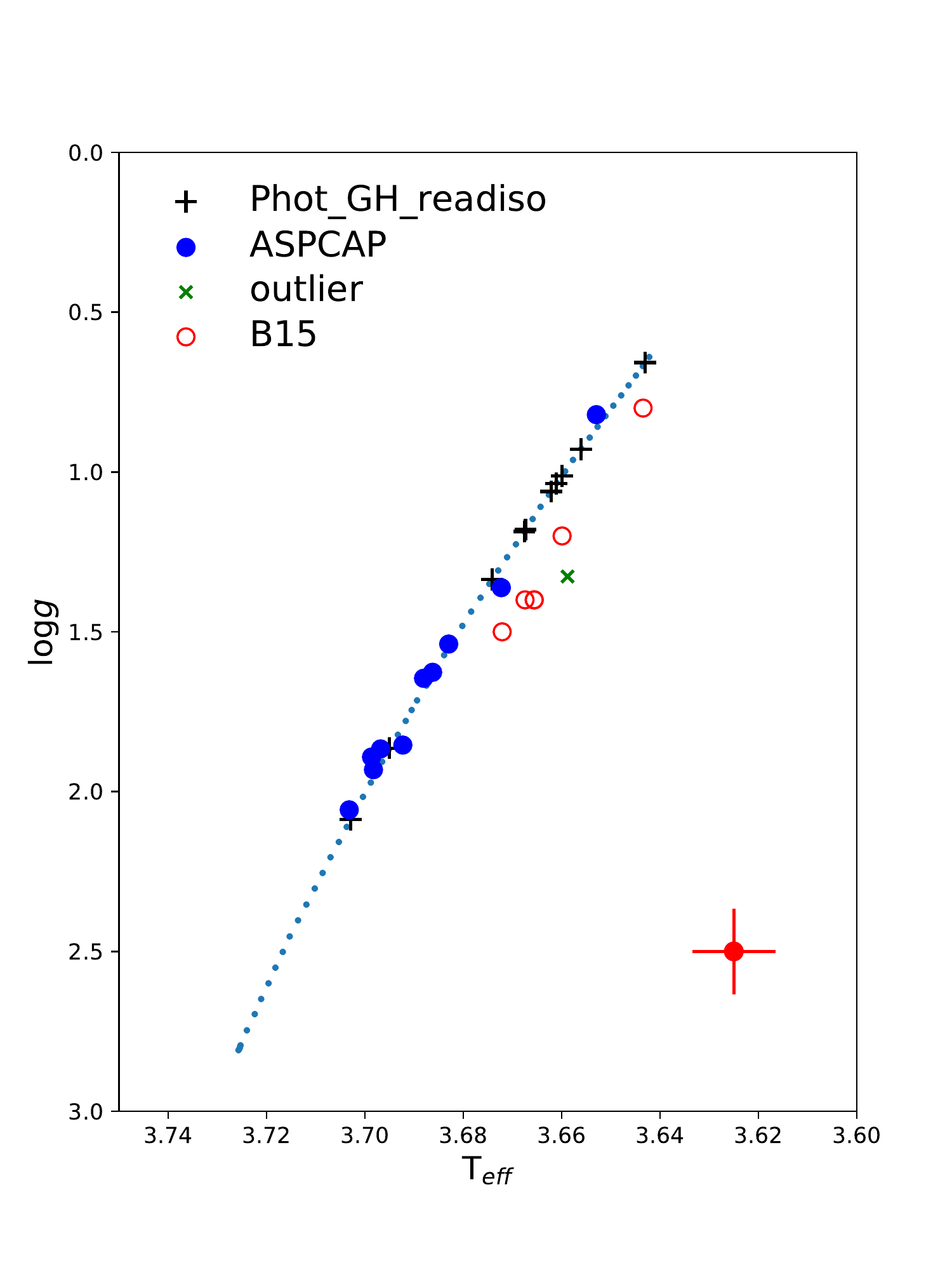} 
\caption{T$_{\rm eff}$ versus $\log g$. The ASPCAP results are labelled with blue dots. The error bars for ASPCAP results are shown in the bottom-right. The photometric stellar parameters using the \citet{GH2009} relations are labelled with plus signs. The B15-derived stellar parameters of the stars in common between our work and B15 are labelled with open circles. The DSED isochrone of 13 Gyr and [Fe/H$]=-2.27$, [$\alpha$/Fe$]=+0.20$ is shown as a dotted line. The cross is the one star that we exclude by CMD and isochrone selection.}\label{fig:iso}
\end{figure}

\section{Chemical abundance derivation}
\label{sect:dev}

Since some of the spectral lines in metal-poor stars become too weak for analysis, stellar parameters (T$_{\rm eff}$, $\log g$, [Fe/H]) derived from $\chi^2$ fitting in ASPCAP may be problematic. \citet{Meszaros2015} suggested that manual analysis is required for stars more metal-poor than [Fe/H]$=-1$. Furthermore, some of the windows for a given element may be dominated by noise when the signal is too weak, which complicates the abundance derivation. Therefore, we  derive chemical abundances manually for our sample of metal-poor stars in NGC 5053. However, note that all of our spectra have S/N$>$75 (Table 2), and most have S/N$>$100.

First, we estimated the stellar parameters using $B,V,I$ \citep{Sarajedini1995} and 2MASS $J,H,K$ photometry (Table \ref{tab:pho}). The photometric T$_{\rm eff}$ is the mean value of the T$_{\rm eff}$ calculated with $B-V$ and $V-K_s$ colors using the equations of \cite{GH2009}. The foreground extinction for each band is given by \citet{Schlafly2011}\footnote{NED. https://ned.ipac.caltech.edu} and is very low ($E(B-V)=0.015$).  
To derive $\log g$, we use the Dartmouth Stellar Evolution Database \citep[DSED,][]{Dotter2008} isochrone of age$=$13 Gyr, [Fe/H$]= -2.27$, [$\alpha$/Fe$]=+0.20$. The $\log g$ for each star is read from the isochrone for its given photometric T$_{\rm eff}$.

Figure \ref{fig:iso} shows the isochrone, stellar parameters (SPs) from ASPCAP and from photometry. We find that (1) the ASPCAP derived SPs agree well with the isochrone; (2) the photometric derived SPs also agree well with the isochrone by definition; (3) the B15 SPs have higher $\log g$ or lower T$_{\rm eff}$ than the isochrone; (4) The number distribution of stars along the isochrone for ASPCAP SPs agrees well with that of CMD in Figure \ref{fig:cmd}, while photometric derived SPs suggest most stars to have low $\log g$ and low T$_{\rm eff}$. 

The iron abundances of NGC 5053 stars have been determined by several optical investigations \citep[e.g., B15, ][]{Sbordone2015}, and the results from the literature are generally around $-2.1$ to $-2.5$ (see B15 for a list of literature results). The value given in the Harris catalog is [Fe/H$]=-2.27$.
After visually inspecting the iron lines in the spectra, we find that the iron lines are hard to detect even with S/N$\sim$100. Given that the derived chemical abundances are affected by the uncertainties of adopted stellar parameters, we decided to derive the elemental abundances using three different sets of stellar parameters:
\begin{enumerate}
\item {\bf SP1}: Photometric derived T$_{\rm eff}$ and $\log g$ from the isochrone, [Fe/H] derived from the iron lines;
\item {\bf SP2}: ASPCAP-calibrated T$_{\rm eff}$ and $\log g$, but assumed [Fe/H$]=-2.27$;
\item {\bf SP3}: B15 derived stellar parameters. 
\end{enumerate}

The micro-turbulence velocity ($v_t$) is calculated using the equation from \citet{Meszaros2015}:
\begin{equation}
v_t=2.24-0.3 \times \log g
\end{equation}


\begin{table*}
\caption{Stellar Parameters of the NGC 5053 Members from ASPCAP.}              
\label{tab:asp}      
\centering                                      
\begin{tabular}{c c c c c c c c c c}         
\hline\hline     
\# & RV & $\delta_{\rm RV}$ & [Fe/H] & $\delta_{\rm [Fe/H]}$ &  T$_{\rm eff}$ &  $\delta_{\rm T_{eff}}$ & $\log g$ & $\delta_{\log g}$&SNR\\
\hline
&\multicolumn{2}{c}{(km/s)}& & &\multicolumn{2}{c}{(K)}& & & \\
\hline
Star1 & 43.81 &  0.17 & -2.28 &  0.02 & 4923.7 &  99.3 &  1.85 &  0.13 & 108 \\
Star2 & 44.56 &  0.06 & -2.36 &  0.02 & 4497.3 &  67.7 &  0.82 &  0.11 & 187 \\
Star3 & 45.19 &  0.06 & -2.24 &  0.02 & 4855.6 &  96.3 &  1.63 &  0.13 & 105 \\
Star4 & 46.02 &  0.15 & -2.31 &  0.02 & 4819.1 &  94.2 &  1.54 &  0.13 & 105 \\
Star5 & 47.12 &  0.10 & -2.52 &  0.03 & 4991.9 & 112.0 &  1.93 &  0.15 & 76 \\
Star6 & 40.68 &  0.13 & -2.35 &  0.02 & 4975.2 &  95.3 &  1.87 &  0.12 & 138 \\
Star7 & 41.97 &  0.18 & -2.25 &  0.02 & 4996.4 & 104.3 &  1.89 &  0.14 & 112 \\
Star8 & 41.86 &  0.38 & -2.48 &  0.03 & 5048.9 & 112.2 &  2.06 &  0.15 & 91 \\
Star9 & 40.77 &  0.15 & -2.44 &  0.02 & 4701.9 &  88.6 &  1.36 &  0.14 & 104 \\
Star10 & 41.81 &  0.10 & -2.35 &  0.02 & 4876.2 &  88.0 &  1.65 &  0.12 & 153 \\
\hline                                             
\end{tabular}

\raggedright{Note: The RVs have been corrected by visually inspection of the strong Mg, Al, and Si spectral lines. See text for more details.}\\
\end{table*}

\begin{table}
\caption{Stellar Parameters 1 (SP1).}              
\label{tab:sp1}      
\centering                                      
\begin{tabular}{c c c c }         
\hline\hline     
\# &  T$_{\rm eff}$   & $\log g$ & [Fe/H] \\
\hline
star1  &  4722.05 &  1.34 &  -2.18\\
star2  &  4395.42 &  0.66 &  -2.19\\
star3  &  4648.89 &  1.18 &  -2.21\\
star4  &  4593.50 &  1.06 &  -2.17\\
star5  &  4954.46 &  1.86 &  -2.20\\
star6  &  4651.53 &  1.19 &  -2.20\\
star7  &  4570.49 &  1.01 &  -2.19\\
star8  &  5045.27 &  2.09 &  -2.17\\
star9  &  4529.76 &  0.93 &  -2.24\\
star10  &  4582.29 &  1.04 &  -2.20\\
\hline                                             
\end{tabular}

\end{table}

\begin{table}
\caption{Stellar Parameters (SP3) and Abundances \\Derived from Optical Spectra (B15).}              
\label{tab:sp3}      
\centering                                      
\begin{tabular}{c c c c c}         
\hline\hline     
\# &  T$_{\rm eff}$   & $\log g$ & [Fe/H] & [Na/Fe]\\
\hline
star2  & 4400 &  0.80 & $-2.43$ & $<0.0$ \\
star3  & 4630 &  1.40 & $-2.48$ & ... \\
star4  & 4700 &  1.51 & $-2.45$ & 0.9 \\
star6  & 4650 &  1.40 & $-2.45$ & ... \\
star7  & 4630 &  1.40 & $-2.47$ & 0.8 \\
star10  & 4570 &  1.20 & $-2.44$ & 1.0 \\
\hline                                             
\end{tabular}

\end{table}

After visually inspecting the APOGEE spectra of the member stars, we find that only Mg, Al, and Si can be reliably derived for most of the stars, due to their extreme metal-poor nature. We use the MARCS/Turbospectrum stellar libraries from \citet{Zamora2015}, where we interpolate the atmospheric model spectra of each star from these model grids. The line list used in this work is the latest internal DR14 atomic/molecular linelist (linelist.20170418).  The wavelengths of the line centers of the Mg, Al, and Si atomic lines are taken from \citet{Souto2016}. We first inspect the strong lines, and apply additional RV shifts on top of the APOGEE RV correction if necessary. Note that APOGEE pipeline determined RV may have a small offset from the true value in the case of weak spectral lines. The line-by-line analysis is done using two codes: (1) A simple python script written by us, based on the 1D LTE spectral synthesis code Turbospectrum \citep{Plez2012}. We call this TBS for short; and (2)  BACCHUS, which is also based on Turbospectrum. We use TBS for stellar parameters SP2 and SP3, and use BACCHUS for stellar parameters SP1.

\begin{figure*}
\centering
\includegraphics [angle=270,width=0.85\textwidth]{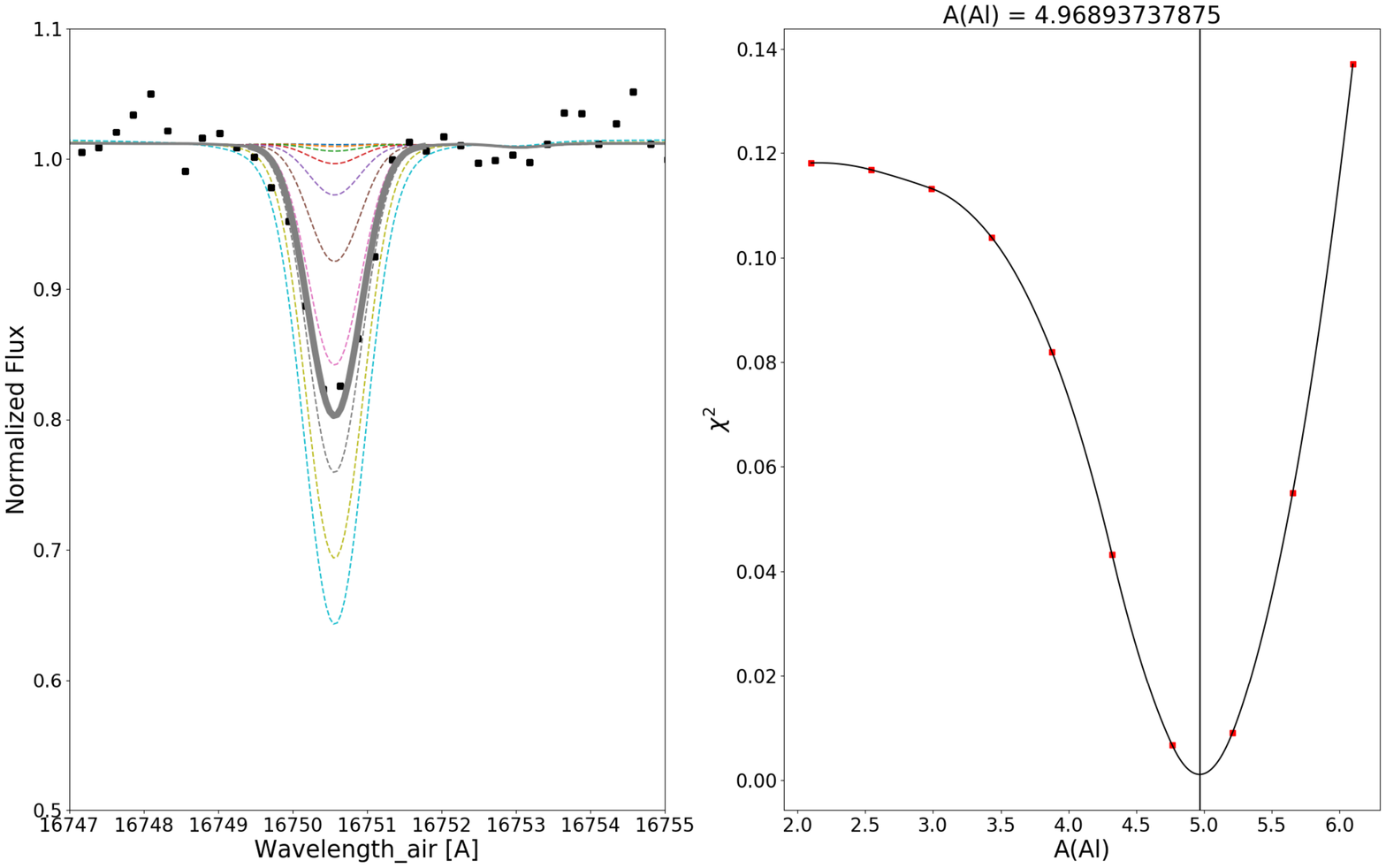} 
\caption{Example of spectral line fitting. In the left panel, the observed spectral line is shown as discrete dots, while models with different Al abundances are shown as lines with different colors. In the right panel, the $\chi^2$ of models with different Al abundances are plotted. A quadratic polynomical interpolation is used to find the $\chi^2$ minimum, as indicated by the vertical line.}\label{fig:fit}
\end{figure*}

We first briefly introduce TBS.
Each spectrum is normalized locally to avoid possible offset in flux, then the $\chi^2$ minimum is searched between the observed and model spectra. An example is given in Figure \ref{fig:fit}. At this stage, visual inspection is required to weed out extremely weak lines and bad fits. Finally, the means of individual measurements are taken as the chemical abundances, and standard deviations as errors (Table \ref{tab:chem}). For BACCHUS, readers are referred to \citet{Masseron2016} for a detailed description. Here we only briefly describe the basic features in BACCHUS. First, a sigma-clipping is applied on the selected continuum points around the targeted line, then a linear fit is computed over the remaining points. Line broadening effects include not only the one caused by microturbulence, but also instrumental broadening. The code can detect significant bad fits, like a sudden drop in the spectrum due to bad pixels in the detectors. Observed spectra and model spectra are compared in four different ways to determine abundances: $\chi^2$ minimization, line intensity, equivalent width, and spectral synthesis. We use $\chi^2$ minimization in this work for consistence between the two codes. Similar to TBS, the means of individual measurements are taken as the chemical abundances, and standard deviations as errors. Between these two codes,
 BACCHUS is faster and also more sophisticated in handling the continuum fitting, line broadening, and other issues, while visual inspection in TBS ensures that our results are not overly influenced by the noise in the spectra.
The measured chemical abundances are given with respect to the solar abundances from \citet{Asplund2005}.

%

\begin{table}
\caption{Manually Derived Chemical Abundances of the NGC 5053 Members.}              
\label{tab:chem}      
\centering                                      
\setlength{\tabcolsep}{2pt} 
\begin{tabular}{c c c c c c c }         
\hline\hline
\multicolumn{7}{|c|}{BACCHUS+SP1}      \\     
\hline     
\# & [Mg/Fe] & $\delta_{\rm [Mg/Fe]}$ & [Al/Fe] & $\delta_{\rm [Al/Fe]}$ &  [Si/Fe] &  $\delta_{\rm [Si/Fe]}$ \\
\hline
star1  &  0.14 &  0.07 &  0.93 &  0.18 &  0.45 &  0.10 \\
star2  &  0.26 &  0.05 &  0.26 &  0.04 &  0.40 &  0.08 \\
star3  &  0.18 &  0.08 &  1.01 &  0.13 &  0.55 &  0.02 \\
star4  &  0.12 &  0.02 &  0.96 &  0.20 &  0.54 &  0.09 \\
star5  &  0.22 &  0.01 &   ... &   ... &  0.51 &  0.10 \\
star6  &  0.12 &  0.03 &  0.83 &  0.07 &  0.51 &  0.03 \\
star7  &  0.21 &  0.07 &  0.79 &  0.16 &  0.46 &  0.05 \\
star8  &  0.30 &  0.06 &   ... &   ... &  0.37 &  0.07 \\
star9  &  0.22 &  0.04 &   ... &   ... &  0.32 &  0.07 \\
star10  &  0.18 &  0.05 &  0.81 &  0.12 &  0.48 &  0.09 \\
\hline                                             
\hline
\multicolumn{7}{|c|}{TBS+SP2}      \\
\hline
\# & [Mg/Fe] & $\delta_{\rm [Mg/Fe]}$ & [Al/Fe] & $\delta_{\rm [Al/Fe]}$ &  [Si/Fe] &  $\delta_{\rm [Si/Fe]}$ \\
\hline
star1  &  0.23 &  0.09 &  0.93 &  0.08 &  0.60 &  0.13 \\
star2  &  0.30 &  0.06 &  0.33 &  0.06 &  0.40 &  0.04 \\
star3  &  0.22 &  0.08 &  1.08 &  0.11 &  0.66 &  0.06 \\
star4  &  0.22 &  0.03 &  0.98 &  0.17 &  0.63 &  0.09 \\
star5  &  0.19 &  0.04 &   ... &   ... &  0.63 &  0.08 \\
star6  &  0.27 &  0.07 &  1.00 &  0.10 &  0.66 &  0.06 \\
star7  &  0.37 &  0.07 &  1.02 &  0.13 &  0.71 &  0.06 \\
star8  &  0.31 &  0.04 &   ... &   ... &  0.40 &  0.04 \\
star9  &  0.27 &  0.08 &   ... &   ... &  0.37 &  0.07 \\
star10  &  0.32 &  0.06 &  0.97 &  0.07 &  0.63 &  0.10 \\
\hline                                             
\hline
\multicolumn{7}{|c|}{TBS+SP3}      \\
\hline
\# & [Mg/Fe] & $\delta_{\rm [Mg/Fe]}$ & [Al/Fe] & $\delta_{\rm [Al/Fe]}$ &  [Si/Fe] &  $\delta_{\rm [Si/Fe]}$ \\
\hline
star2  &  0.40 &  0.05 &  0.40 &  0.06 &  0.52 &  0.09 \\
star3  &  0.32 &  0.05 &  1.15 &  0.12 &  0.74 &  0.04 \\
star4  &  0.33 &  0.02 &  1.09 &  0.17 &  0.75 &  0.05 \\
star6  &  0.30 &  0.06 &  1.00 &  0.11 &  0.66 &  0.05 \\
star7  &  0.41 &  0.08 &  1.02 &  0.14 &  0.72 &  0.05 \\
star10  &  0.35 &  0.06 &  0.98 &  0.10 &  0.64 &  0.09 \\
\hline
\end{tabular}
\end{table}

\begin{figure*}
\centering
\includegraphics [width=0.95\textwidth]{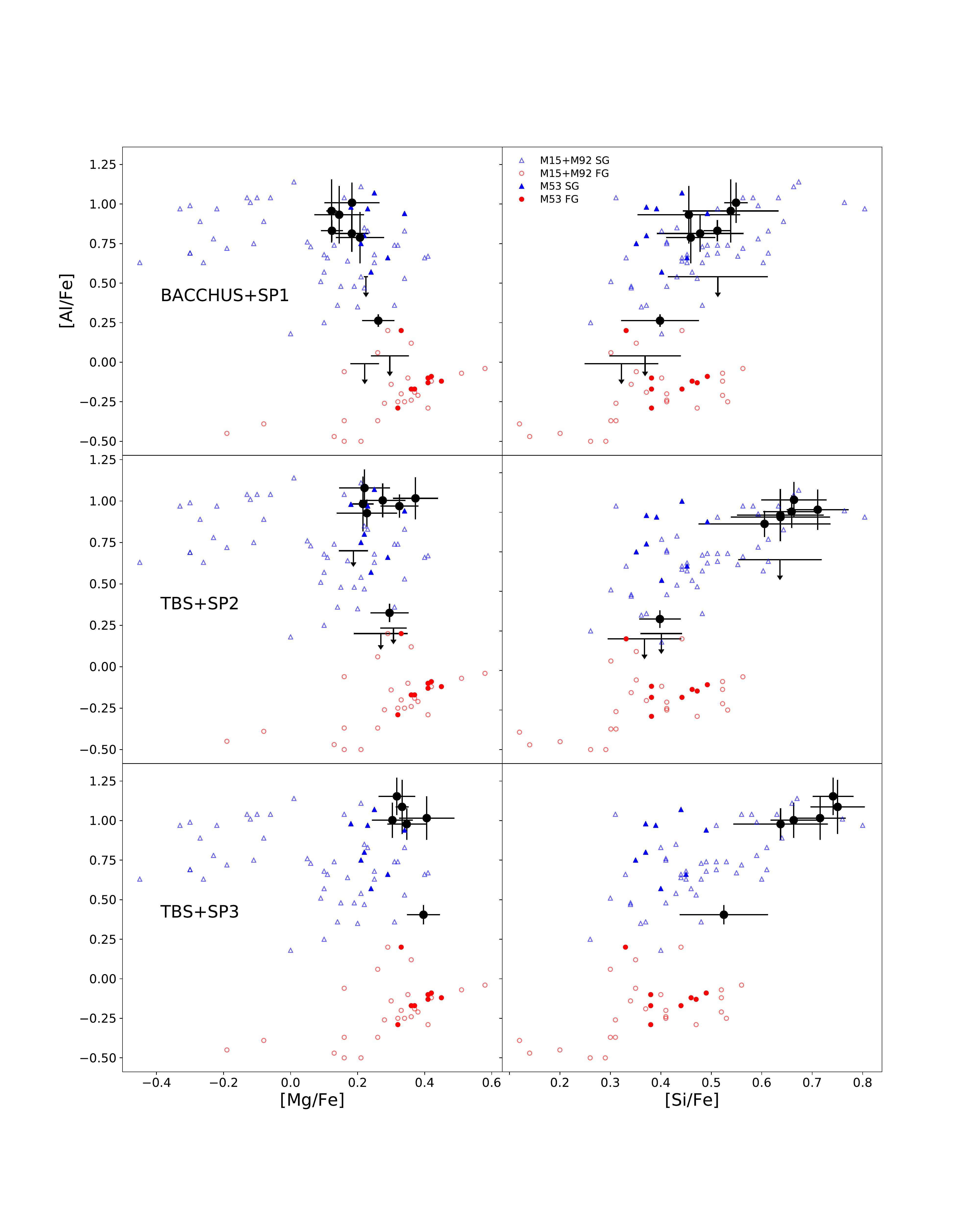} 
\caption{ $Left~column$: [Al/Fe] versus [Mg/Fe]. $Right~column$: [Al/Fe] versue [Si/Fe]. $Top~row$: SP1 and BACCHUS. $Middle~row$: SP2 and TBS. $Bottom~row$: SP3 and TBS. The NGC 5053 stars are black dots. The two generations of stars in the metal-poor GCs from \citet{Meszaros2015} are labelled as red circles (FG) and blue triangles (SG), with M15 and M92 as open symbols, and M53 as filled ones. 
Three stars with estimated Al upper limits are indicated by arrows.}\label{fig:BAC}
\end{figure*}

%

\section{Results}
\label{sect:chem}

\subsection{Multiple Populations in NGC 5053}

We derive the iron abundances using BACCHUS, listed in Table \ref{tab:sp1}. The mean iron abundance is $-2.20$ with a standard deviation of 0.02 dex. Our measurements basically agree with the iron abundance given by the Harris catalog ([Fe/H$]=-2.27$).  The iron lines in NIR are generally weaker and less numerous compared to those in the optical; therefore an offset of $\sim$0.1 dex between the iron abundances derived from the NIR and from the optical is acceptable.

Visual inspection suggests that for three stars (Star5, Star8, Star9) we have no reliable Al measurements because the lines are too weak. However, we managed to inspect these lines individually to estimate upper limits. We plot the measured abundances and upper limits for the three different sets of stellar parameters in Figures \ref{fig:BAC}. To study the multiple populations in metal-poor GCs in context, we also plot three metal-poor GCs (M15, M92, M53) from \citet{Meszaros2015} that have [Fe/H$]<-2$. The first and second generations that were defined in \citet{Meszaros2015} are distinguished by red circles and blue triangles, respectively. Interestingly, M53 ([Fe/H$]=-2.10$), which is a close companion of NGC 5053 in the sky, was also included in that work. We label the M53 stars with filled symbols. 
A clear Al variation as large as 0.8 dex is found in NGC 5053, and we can easily separate two groups of stars with different Al abundances. This is true for all three sets of SPs. We note that six stars are in common between our sample and B15 (Table \ref{tab:pho}), and four of them have [Na/Fe] measurements from optical spectra (Table \ref{tab:sp3}). The only first-generation (FG) star (Star2) in common shows a low [Na/Fe] ($< 0$), and three second-generation (SG) stars in common show [Na/Fe$]\sim0.9$. Thus, our stellar generation division by Al is supported by the Na abundances from B15. The two stellar generations of NGC 5053 also fit well into the two stellar generations defined by the other metal-poor GCs, though we note that three FG stars in this work (Star2, Star8, Star9) lie at the limit that separates the two stellar generations, although two of these stars only have Al upper limits from our data and therefore could actually lie closer to the FG star region. Meanwhile, Mg shows no strong variation between the two stellar populations. In all three sets of measurements, we do not find stars with [Mg/Fe$]<0.0$. The Mg-Al anti-correlation of NGC 5053 does not resemble that of M92 or M15 ([Fe/H$]\sim -2.3$, open symbols), but resembles that of M53 ([Fe/H$]\sim -2.1$, filled symbols); in the sense that no strong Mg depletion is seen in NGC 5053, even in SG stars. 

All three sets of chemical measurements suggest a substantial Si abundance variation ($\gtrsim$0.25 dex), which leads to a Si-Al correlation. Note that the Si separation between the two stellar generations is smaller than the Al separation. In terms of Si distribution, NGC 5053 stars  resemble those of M92 or M15, but not of M53. The different behaviors of Mg and Si among metal-poor GCs suggest that metallicity is not the only parameter that regulates the MP phenomenon (see Section \ref{sect:mp}).

\begin{figure}
\centering
\includegraphics [width=0.39\textwidth]{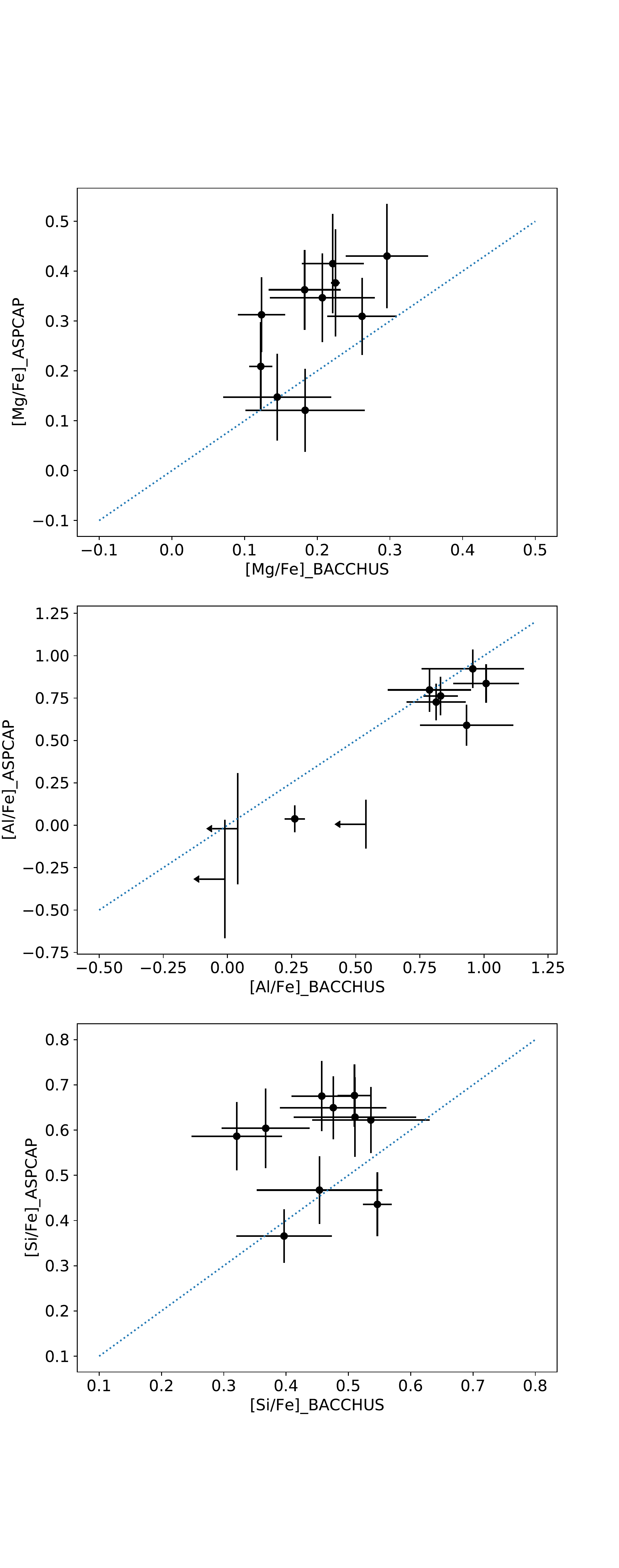} 
\caption{Comparison between the BACCHUS$+$SP1 and ASPCAP pipeline results. The dotted lines represent the one-to-one line.}\label{fig:comp}
\end{figure}

\begin{figure}
\centering
\includegraphics [width=0.39\textwidth]{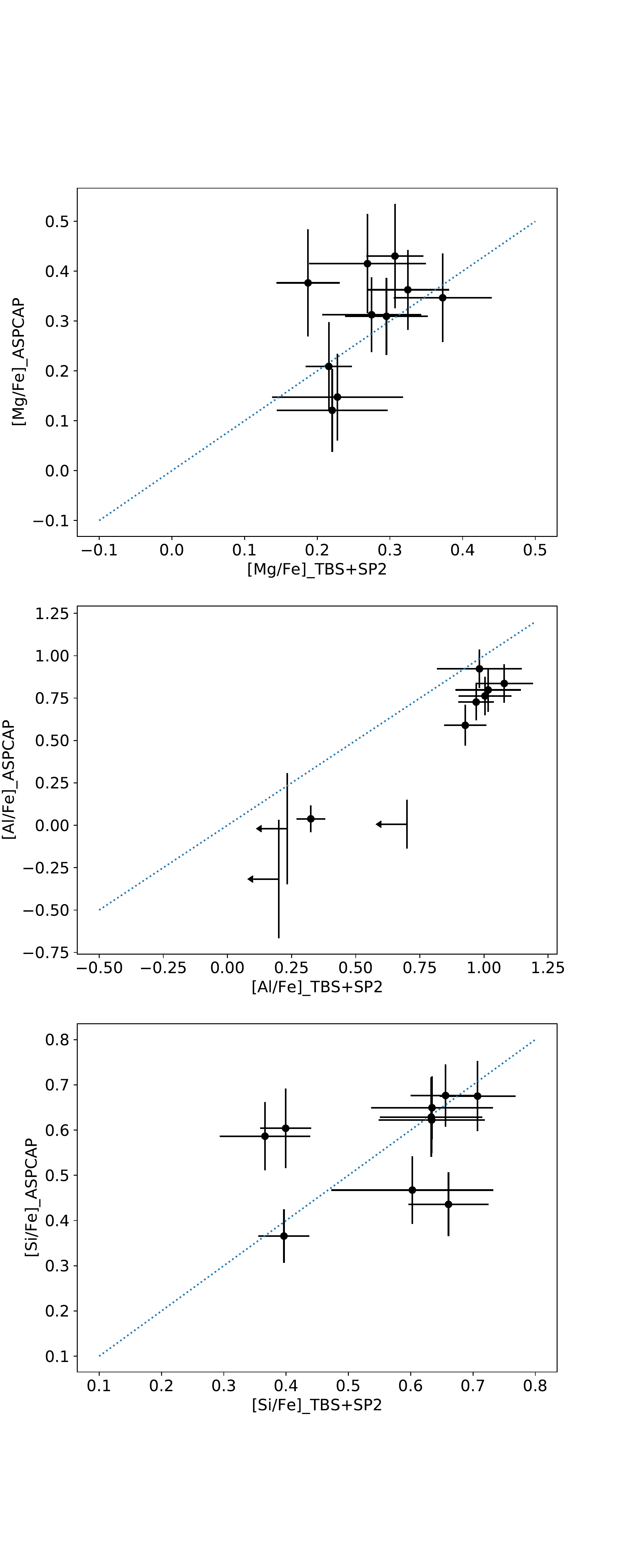} 
\caption{Comparison between the TBS$+$SP2 and ASPCAP pipeline results. The dotted lines represent the one-to-one line.}\label{fig:comp2}
\end{figure}

\subsection{Robustness of the Measurements}
\label{sect:rob}

In this work, we derive elemental abundances adopting different sets of stellar parameters and different codes. SP3 (B15) usually yields higher elemental abundances due to the assumed low metallicities ([Fe/H$]\sim-2.45$). Conversely, SP1 usually yields lower elemental abundances: for example, the [Mg/Fe] abundances are lower in the SG stars, where an Al-Mg anti-correlation is barely seen.
Bearing these uncertainties in mind, our conclusions, namely, that there is a range of Al and Si abundances in NGC 5053 stars, and that those are correlated, are insensitive to the choice of stellar parameters adopted in the analysis. 

Figure \ref{fig:comp} compares the elemental abundances derived by ASPCAP and the BACCHUS$+$SP1 method. ASPCAP derived [Mg/Fe] and [Si/Fe] are higher than those of the BACCHUS results, partially due to the lower [Fe/H] from the ASPCAP pipeline ([Fe/H$]=-2.36\pm0.09$). There is no one-to-one correlation between the two sets of results. On the other hand, the [Al/Fe] abundances are similar between the two methods, though we note that the upper limits of the chemical abundances can only be given after visual inspection under the circumstance of very weak lines. We find a similar situation if we compare the elemental abundances derived by ASPCAP and the TBS$+$SP2 method (Figure \ref{fig:comp2}). Though the systematic offset between the Mg (and Si) abundances from these two methods is smaller, their comparison does not show a one-to-one correlation. However, the [Al/Fe] abundances derived from the two methods follow a one-to one correlation, with possible systematic offset.
Therefore, we emphasize that it is necessary to derive chemical abundances manually in the case of very metal-poor stars, especially for those stars with [Fe/H$]<-2$. We note that the ASPCAP performance at extremely low [Fe/H] has been a known issue in the ASPCAP team.

\section{Discussion}
\label{sect:dis}

\subsection{Is NGC 5053 Associated with Sgr?}

\begin{figure*}
\centering
\includegraphics [width=0.85\textwidth]{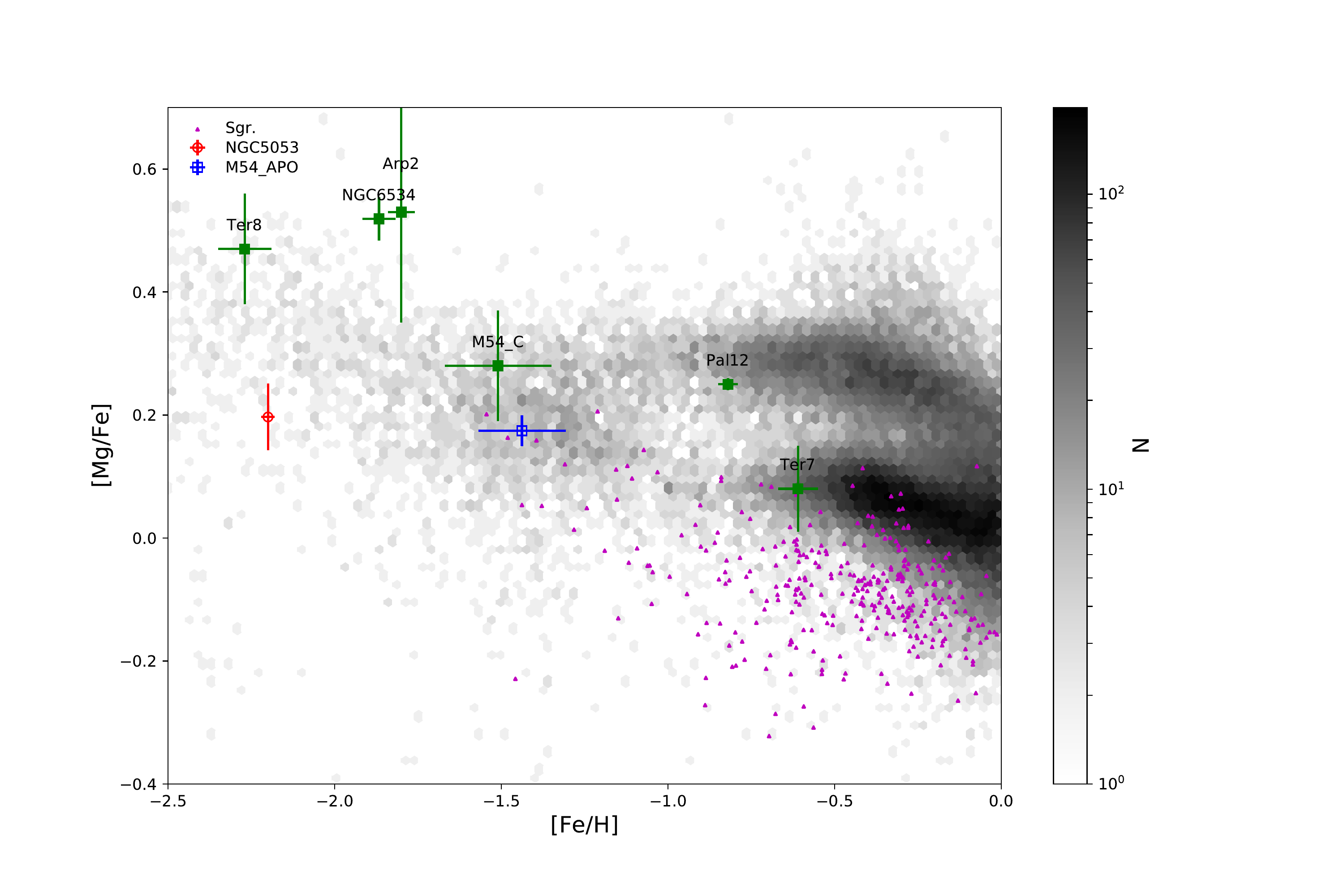} 
\caption{[Mg/Fe] versus [Fe/H]. The Sgr stars selected by \citet{Hasselquist2017} are plotted as magenta triangles. The red open circle represents the average [Mg/Fe] and [Fe/H] abundances for NGC 5053 (BACCHUS$+$SP1). The blue open square represents the average [Mg/Fe] and [Fe/H] abundances of the M54 members.  GCs with possible Sgr origin are shown as green filled squares, and their GC names are labelled.
We select a sample of stars from the APOGEE survey to visualize the thin disk, thick disk, and halo (see text). The grey scale map indicates the number density of these selected stars on a logarithmic scale.}\label{fig:mgfe}
\end{figure*}

Using the newly obtained, high-precision proper motion of NGC 5053 and our software GravPot16, we have shown that the orbit of NGC 5053 is significantly different from that of Sgr. Our simulation results strongly suggest against a connection between NGC 5053 and Sgr.

In terms of chemistry, we first take a look at [Mg/Fe] vs. [Fe/H], since the Mg abundances of NGC 5053 are newly obtained in this work, and \citet{Hasselquist2017} have shown that the Milky Way (MW) thin disk, thick disk, and Sgr core stars are distinguishable in this parameter space. To visualize the MW thin disk, thick disk and halo, we first select giants from APOGEE DR14 with the following criteria: (1) $3500 < \rm T_{eff} <5000$ K; (2) $\log g <2.0$; (3) SNR$>$70; (4) [Fe/H] and [Mg/Fe] measurements are available. The number density of these stars on a logarithmic scale is indicated by a grey scale map in Figure \ref{fig:mgfe}, where high number density regions are darker. The stars selected by Hasselquist et al. are plotted as magenta triangles. Besides NGC 5053, we also try to identify other possible Sgr-associated GCs in APOGEE. We preliminarily identify five stars from M54, using a similar method to that described in Section \ref{sect:data}. We have checked the strong Mg lines manually, and find that the ASPCAP fits are reasonable. Therefore, we use the ASPCAP values instead of performing another manual analysis. We note that more observations of M54 stars have been scheduled by the APOGEE team; a careful analysis of M54 stars will be left to a future publication.  We also gather data for other possible Sgr-associated GCs from literature: Terzan 8 \citep{Carrettater8}, M54 \citep{Carrettam54}, Palomar 12 \citep{Cohen2004}, Terzan 7 \citep{Sbordone2007}, Arp 2 \citep{Mottini2008}, and NGC 6534 \citep{Carretta2017}. The literature [Fe/H] and [Mg/Fe] are shown as green squares in Figure \ref{fig:mgfe} with associated measurement errors. Note that [Fe/H] and [Mg/Fe] measured by \citet{Carrettam54} and APOGEE are consistent within their uncertainty ranges.

We see that the MW thin disk, thick disk, and the Sgr core sequences are indeed well separated at [Fe/H$]>-1.0$ (Figure \ref{fig:mgfe}). At lower metallicity, MW thin disk and thick disk do not have much stars in this region. The stars between $-2.0<[$Fe/H$]<-0.9$ are suggested to be halo stars with two different chemical and dynamical properties \citep{Hayes2018, FA2018}. 
We run out of Sgr stars at [Fe/H$]<-1.7$, and the MW halo sequence becomes wide and sparse, so it is difficult to estimate whether NGC 5053 is chemically similar to Sgr or MW halo using only [Mg/Fe] and [Fe/H]. On the other hand, the possible Sgr-associated GCs loosely form a sequence in the low metallicity regime (green squares).
The average Mg abundance of NGC 5053 appears lower than the barely-seen MW halo sequence and its value ($\sim0.2$ dex) is lower than that of Terzan 8 ([Mg/Fe$]=0.47$ dex), which has similar metallicity as NGC 5053 ([Fe/H$]=-2.27$, \citealt{Carrettater8}). This indicates that NGC 5053 may be chemically different than other Sgr GCs,
though we caution that the Mg abundances of SG stars may be slightly affected by the Mg-Al reaction chains and that the Sgr GC trend is highly dependent on the three most metal-poor GCs.  Meanwhile, studies of metals (e.g., Ni), and heavy elements (e.g., Y and Ba) in NGC 5053, have not shown conclusive evidence of a Sgr connection \citep{Sbordone2015}. Unfortunately, Ni and s-process elements (e.g., Nd and Ce) do not show any measurable lines in our APOGEE spectra.

To summarize, we find tentative chemical evidence to dismiss a connection between NGC 5053 and Sgr, and our simulated orbit of NGC 5053 also argues against such a connection.

\subsection{Multiple Populations in the Metal-Poor Regime}
\label{sect:mp}
\begin{table*}
\caption{Basic Parameters for GCs in this Paper.}              
\label{tab:gc}      
\centering                                      
\begin{tabular}{c c c c c c}         
\hline\hline     
ID &Name& [Fe/H]$^{a}$ &$\delta_{\rm [Fe/H]}^{a}$ & $\log \rm Mass1^{b}$ (M$_{\odot}$) &Mass2$^{c}$ (M$_{\odot}$)\\
\hline
NGC 5053 &     & $-2.30$ & 0.08 & $4.81\pm0.01$ & $5.37\pm 1.32\times 10^4$ \\
NGC 7078 & M15 & $-2.33$ & 0.02& ... & $5.01\pm 0.06\times 10^5$ \\
NGC 6341 & M92 & $-2.35$ & 0.05& $5.45\pm0.01$ & $3.05\pm 0.04\times 10^5$ \\
NGC 5024 & M53 & $-2.06$ &0.09 & $5.66\pm0.01$ & $3.83\pm 0.51\times 10^5$ \\
NGC 2808 &        & $-1.18$ &0.04&  $5.93 \pm0.01$ &$8.29\pm 0.06\times 10^5$ \\
NGC 6388 &       & $-0.45$  & 0.04&  $6.05\pm0.05$  & $1.24\pm 0.01\times 10^6$ \\
NGC 6752 &       & $-1.55$ & 0.01  &   ... & $2.34\pm 0.04\times 10^5$\\
\hline                                             
\end{tabular}

\hspace{1.4in}\raggedright{$^a$ From \citet{Carretta2009iron}.}\\
\hspace{1.4in}\raggedright{$^b$ From \citet{McLaughlin2005}.}\\
\hspace{1.4in}\raggedright{$^c$ From \cite{Baumgardt2017}.}\\
\end{table*}

After carefully inspecting three elements involved in the Mg-Al nuclear chain for four metal-poor GCs with different GC cluster masses (Table \ref{tab:gc}), we see interesting behaviors in their abundance variations. The Mg-Al nuclear chain and the associated  Si leakage are supposed to be activated at temperature above 90 MK \citep{Prantzos2007, Ventura2012, DAntona2016}, with stronger Si leakage at higher temperature. Strong Mg depletion is observed in the two more metal-poor ([Fe/H$]\sim -2.3$) and massive GCs M92 and M15, but not in the similar metallicity but less massive GC NGC 5053, or in the less metal-poor ([Fe/H$]\sim -2.1$) but massive GC M53. On the other hand, a Si spread is found for the three most metal-poor GCs M92, M15, and NGC 5053, regardless of their cluster masses, while a Si spread is not found in the less metal-poor, massive GC M53. Therefore, the correlations among Mg, Al, and Si for MPs are not only affected by metallicity, other parameters may also play important roles. Cluster mass has been suggested to be another parameter that affects the elemental abundances of MPs \citep[e.g., ][]{Carretta2010b, Pancino2017}. In our work, we notice that NGC 5053 has a lower cluster mass (Table \ref{tab:gc}) and exhibits less prominent Al-Mg anti-correlation compared to M15 and M92. This qualitatively agrees with the trend given in \citet{Pancino2017}, where the standard deviation of the [Al/Mg] distribution and its maximum variation of low mass GCs tend to be smaller in a given [Fe/H].


\cite{Carretta2009b} suggested that Si variation is limited to massive or metal-poor GCs, based on the four GCs that show Si variation in their observations. This observational evidence agrees with our understanding of the nucleosynthesis reaction chains that are responsible for MPs, since more massive stars or more metal-poor stars tend to have higher reaction temperature, which is required for the Mg-Al nuclear chain reaction and the associated Si leakage \citep{Arnould1999}. Including the GCs that show Si variation in their sample, NGC 2808, NGC 6388, NGC 6752, and NGC 7078 (M15), we found that NGC 5053 has the lowest cluster mass (Table \ref{tab:gc}). Depending on the mass estimation methods \citep{McLaughlin2005, Baumgardt2017}, the masses of the clusters might be slightly different. However, we notice that the mass of NGC 5053 is constantly about one order of magnitude lower than other GCs. Therefore, this cluster likely has the lowest cluster mass among the GCs that have been identified to exhibit an observable Si spread until now.

Nevertheless, we caution that our sample size is limited. Ten stars are a reasonably statistically significant sample, but there is still a small chance that we may be missing the strong Mg-depleted stars. Besides the ten stars that we study in this work, the CMD of NGC 5053 (Figure \ref{fig:cmd}) indicates that another $\sim10$ bright, RV-identified cluster members (B15) may be observable with the APOGEE-2 spectrograph (without considering fiber collision). Since the Mg, Al, and Si lines are more difficult to be detected in the optical for metal-poor stars \citep{Sbordone2015}, NIR high-resolution spectroscopy, e.g., APOGEE-2, is more appropriate. 

\section{Summary}

Driven by the desire to understand multiple populations in the oldest, most metal-poor stellar systems in the nearby Universe, we have carefully studied the orbit and chemical abundances of one of the most metal-poor GCs, NGC 5053. Given the similar location and RV between NGC 5053 and one of the Sgr arms, NGC 5053 has been speculated to be associated with Sgr. However, using our dynamical code GravPot16, we argue that a physical connection between Sgr and NGC 5053 is unlikely. 

We have identified 10 cluster members in the APOGEE DR14 using their spatial location, RV, [Fe/H], and CMD location. We detected strong Mg, Al, and Si spectral lines in the APOGEE spectra, which are otherwise difficult to see in optical spectra. We manually analyzed the spectra with three different combinations of stellar parameters and codes. We proved that the large Al variation and the substantial Si spread are independent of the adopted measured method. Along with NGC 5053, metal-poor GCs exhibit different Mg, Al, and Si variations, which agrees with the claim that metallicity is not the only parameter that is fundamental to the MP phenomenon. Parameters such as cluster mass, must also play important roles. Interestingly, NGC 5053 has the lowest cluster mass among all the GCs that are currently found to exhibit Si variations. This may be the lower mass limit to generate an observable Si spread.



\section{acknowledgments}

We thank the anonymous referee for insightful comments. 
BT, JGFT, DG, and SV gratefully acknowledge support from the Chilean BASAL Centro de Excelencia en Astrof\'{i}sica y Tecnolog\'{i}as Afines (CATA) grant PFB-06/2007. 
BT acknowledges support from the one-hundred-talent project of Sun Yat-Sen University.
SV gratefully acknowledges the support provided by Fondecyt reg. n. 1170518.
OZ, TM, DAGH, and FD acknowledge support provided by the Spanish Ministry of Economy and  Competitiveness  (MINECO)  under  grant AYA-2014-58082-P. DAGH was also funded by  the  Ram\'on  y  Cajal  fellowship  number RYC-2013-14182.
SzM has been supported by the Premium Postdoctoral
Research Program of the Hungarian Academy of Sciences, and by the Hungarian
NKFI Grants K-119517 of the Hungarian National Research, Development and Innovation Office.
TCB acknowledges partial support from grant PHY 14-30152; Physics
Frontier Center/JINA Center for the Evolution of the Elements
(JINA-CEE), awarded by the US National Science Foundation.
STS acknowledges support by NASA through grants GO-14235 from the Space Telescope Science Institute (STScI), which is operated by the Association of Universities for Research in Astronomy (AURA), Inc., under NASA contract NAS5-26555.
Funding for the GravPot16 software has been provided by the Centre national d'etudes spatiale (CNES) through grant 0101973 and UTINAM Institute of the Universit\'e de Franche-Comt\'e, supported by the R\'egion de Franche-Comt\'e and Institut des Sciences de l'Univers (INSU).
Monte Carlo simulations have been executed on computers from the Utinam Institute of the Universit\'e de Franche-Comt\'e, supported by the R\'egion de Franche-Comt\'e and Institut des Sciences de l'Univers (INSU).

Funding for the Sloan Digital Sky Survey IV has been provided by the
Alfred P. Sloan Foundation, the U.S. Department of Energy Office of
Science, and the Participating Institutions. SDSS- IV acknowledges
support and resources from the Center for High-Performance Computing at
the University of Utah. The SDSS web site is www.sdss.org.

SDSS-IV is managed by the Astrophysical Research Consortium for the Participating Institutions of the SDSS Collaboration including the Brazilian Participation Group, the Carnegie Institution for Science, Carnegie Mellon University, the Chilean Participation Group, the French Participation Group, Harvard-Smithsonian Center for Astrophysics, Instituto de Astrof\`{i}sica de Canarias, The Johns Hopkins University, Kavli Institute for the Physics and Mathematics of the Universe (IPMU) / University of Tokyo, Lawrence Berkeley National Laboratory, Leibniz Institut f\"{u}r Astrophysik Potsdam (AIP), Max-Planck-Institut f\"{u}r Astronomie (MPIA Heidelberg), Max-Planck-Institut f\"{u}r Astrophysik (MPA Garching), Max-Planck-Institut f\"{u}r Extraterrestrische Physik (MPE), National Astronomical Observatory of China, New Mexico State University, New York University, University of Notre Dame, Observat\'{o}rio Nacional / MCTI, The Ohio State University, Pennsylvania State University, Shanghai Astronomical Observatory, United Kingdom Participation Group, Universidad Nacional Aut\'{o}noma de M\'{e}xico, University of Arizona, University of Colorado Boulder, University of Oxford, University of Portsmouth, University of Utah, University of Virginia, University of Washington, University of Wisconsin, Vanderbilt University, and Yale University.

\appendix
\section{Galactic Model}

For the Galactic model we employ a 3D steady-state gravitational potential of the Galaxy, assumed to be made up of the superposition of many composite stellar populations belonging to the thin disc, the thick disk, the Interstellar Matter (ISM), the stellar and dark matter halo, and a rotating Galactic bar component, that fit fairly well the structural and dynamical parameters of the Milky Way. The mass density distribution of each stellar and non-stellar component is described in detail in \citet{Robin2003, Robin2012, Robin2014} and the consequent semianalytic Milky Way potential shape will be fully described in a forthcoming paper (Fern\'andez-Trincado et al. 2017, in preparation). Here, we present a summary of the main structural parameters of our 3D steady-state potential model; which have been adopted in a score of papers \citep[see e.g.,][]{Fernandez-Trincado2016, FT2017c, Fernandez-Trincado2017a, Fernandez-Trincado2017b, Abolfathi2017, Recio-Blanco2017, Anders2017, Libralato2018}.

The model is made up of the sum of an axisymmetric background potential and a non-axisymmetric one. The axisymmetric potential is dominated by the superposition of homogenous oblate spheroids, which approximate the density models of \citet{Einasto1979} thin disks \citep[see e.g.,][]{Robin2003}, while the ISM and the thick disks are the analytic expressions as introduced in \citet{Smith2015}, which approximate the density profiles in \citet{Robin2003, Robin2014}.  The non-axisymmetric part includes a rotating bar, where the corresponding potential model consists of triaxial inhomogeneous ellipsoids similar to \citet{Pichardo2004}, which approximate the dimensions and density model of \citet{Robin2012}. In this work, the overall Galactic potential has been scaled to the Sun's Galactocentric distance, 8 kpc, and the local rotation velocity of 244.5 km s$^{-1}$ \citep[e.g.,][]{Sofue2015}. The adopted values for the Sun's orbital velocity vector  are [$U_{\odot}, V_{\odot}, W_{\odot}$] $=$ [13.12$\pm$1.47, 0.92$\pm$0.29, 7.03$\pm$0.18] km s$^{-1}$ \citep[e.g.,][]{Robin2017}. For reference, we have adopted a righthanded coordinate system for $U, V$ and $W$, so that they are positive in the directions of the Galactic center, Galactic rotation, and North Galactic Pole, respectively.
 
\begin{figure*}
	\centering
	\includegraphics [angle=270,width=1.0\textwidth]{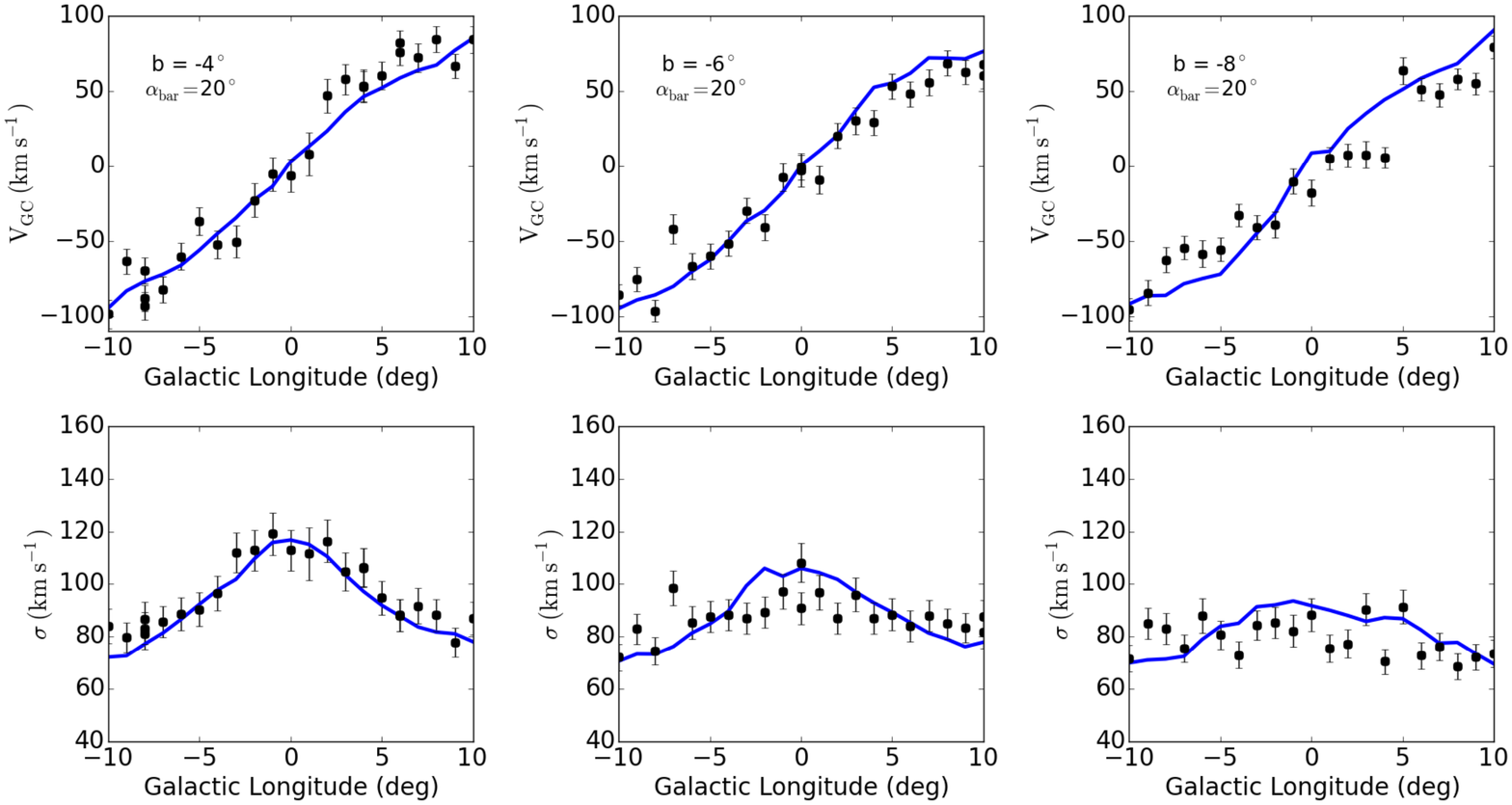} 
	\caption{Rotation curve (blue) of a boxy bulge from 3D test particle simulations evolved in 3D steady-state potential model (\textit{GravPot16}) compared to BRAVA (black) data \citep{Howard2008, Kunder2012}. Taken from Figure 5.12 in \citet{Fernandez-Trincado2017Thesis}}
	\label{Figure1}
\end{figure*}

For the structural parameters of the bar potential we adopted the dynamical constraints from 3D test particle simulations evolved in the 3D steady-state Milky Way potential above mentioned \citep[][]{Fernandez-Trincado2017Thesis, Fernandez-Trincado2017b}, which is able to reproduce the velocity rotation curve of stars in the bulge at different Galactic longitude and latitudes, as illustrated in Figure \ref{Figure1}. For our computations, we have considered the total mass for the bar, 1.1$\times$10$^{10}$ M$_{\odot}$, the present-day orientation of the major axis of the Galactic bar, $\phi=20^{\circ}$, and a pattern speed, $\Omega_{\rm B}= 35$ km s$^{-1}$ kpc$^{-1}$. These values are consistent with the recent estimates from literature \citep[see][for instance]{Portail2015}.

\bibliographystyle{aasjournal}
\bibliography{gc,apo,5053}

\begin{thebibliography}{}
\expandafter\ifx\csname natexlab\endcsname\relax\def\natexlab#1{#1}\fi

\bibitem[{{Abolfathi} {et~al.}(2017){Abolfathi}, {Aguado}, {Aguilar}, {Allende
  Prieto}, {Almeida}, {Tasnim Ananna}, {Anders}, {Anderson}, {Andrews},
  {Anguiano}, \& et~al.}]{Abolfathi2017}
{Abolfathi}, B., {Aguado}, D.~S., {Aguilar}, G., {et~al.} 2017, ArXiv e-prints,
  arXiv:1707.09322

\bibitem[{{Anders} {et~al.}(2017){Anders}, {Queiroz}, {Chiappini}, {Santiago},
  {Fern{\'a}ndez-Trincado}, {Meza}, \& {the SDSS-IV/APOGEE
  Collaboration}}]{Anders2017}
{Anders}, F., {Queiroz}, A.~B., {Chiappini}, C., {et~al.} 2017, ArXiv e-prints,
  arXiv:1708.09319

\bibitem[{{Arellano Ferro} {et~al.}(2010){Arellano Ferro}, {Giridhar}, \&
  {Bramich}}]{AF2010}
{Arellano Ferro}, A., {Giridhar}, S., \& {Bramich}, D.~M. 2010, \mnras, 402,
  226

\bibitem[{{Arnould} {et~al.}(1999){Arnould}, {Goriely}, \&
  {Jorissen}}]{Arnould1999}
{Arnould}, M., {Goriely}, S., \& {Jorissen}, A. 1999, \aap, 347, 572

\bibitem[{{Asplund} {et~al.}(2005){Asplund}, {Grevesse}, \&
  {Sauval}}]{Asplund2005}
{Asplund}, M., {Grevesse}, N., \& {Sauval}, A.~J. 2005, in Astronomical Society
  of the Pacific Conference Series, Vol. 336, Cosmic Abundances as Records of
  Stellar Evolution and Nucleosynthesis, ed. T.~G. {Barnes}, III \& F.~N.
  {Bash}, 25

\bibitem[{{Baumgardt}(2017)}]{Baumgardt2017}
{Baumgardt}, H. 2017, \mnras, 464, 2174

\bibitem[{{Bellazzini} {et~al.}(2003){Bellazzini}, {Ferraro}, \&
  {Ibata}}]{Bellazzini2003}
{Bellazzini}, M., {Ferraro}, F.~R., \& {Ibata}, R. 2003, \aj, 125, 188

\bibitem[{{Boberg} {et~al.}(2015){Boberg}, {Friel}, \&
  {Vesperini}}]{Boberg2015}
{Boberg}, O.~M., {Friel}, E.~D., \& {Vesperini}, E. 2015, \apj, 804, 109

\bibitem[{{Carretta} {et~al.}(2009{\natexlab{a}}){Carretta}, {Bragaglia},
  {Gratton}, {D'Orazi}, \& {Lucatello}}]{Carretta2009iron}
{Carretta}, E., {Bragaglia}, A., {Gratton}, R., {D'Orazi}, V., \& {Lucatello},
  S. 2009{\natexlab{a}}, \aap, 508, 695

\bibitem[{{Carretta} {et~al.}(2009{\natexlab{b}}){Carretta}, {Bragaglia},
  {Gratton}, \& {Lucatello}}]{Carretta2009b}
{Carretta}, E., {Bragaglia}, A., {Gratton}, R., \& {Lucatello}, S.
  2009{\natexlab{b}}, \aap, 505, 139

\bibitem[{{Carretta} {et~al.}(2014){Carretta}, {Bragaglia}, {Gratton},
  {D'Orazi}, {Lucatello}, \& {Sollima}}]{Carrettater8}
{Carretta}, E., {Bragaglia}, A., {Gratton}, R.~G., {et~al.} 2014, \aap, 561,
  A87

\bibitem[{{Carretta} {et~al.}(2010{\natexlab{a}}){Carretta}, {Bragaglia},
  {Gratton}, {Recio-Blanco}, {Lucatello}, {D'Orazi}, \&
  {Cassisi}}]{Carretta2010b}
---. 2010{\natexlab{a}}, \aap, 516, A55

\bibitem[{{Carretta} {et~al.}(2017){Carretta}, {Bragaglia}, {Lucatello},
  {D'Orazi}, {Gratton}, {Donati}, {Sollima}, \& {Sneden}}]{Carretta2017}
{Carretta}, E., {Bragaglia}, A., {Lucatello}, S., {et~al.} 2017, \aap, 600,
  A118

\bibitem[{{Carretta} {et~al.}(2010{\natexlab{b}}){Carretta}, {Bragaglia},
  {Gratton}, {Lucatello}, {Bellazzini}, {Catanzaro}, {Leone}, {Momany},
  {Piotto}, \& {D'Orazi}}]{Carrettam54}
{Carretta}, E., {Bragaglia}, A., {Gratton}, R.~G., {et~al.} 2010{\natexlab{b}},
  \aap, 520, A95

\bibitem[{{Chun} {et~al.}(2010){Chun}, {Kim}, {Sohn}, {Park}, {Han}, {Kim},
  {Lee}, {Lee}, {Lee}, \& {Sohn}}]{Chun2010}
{Chun}, S.-H., {Kim}, J.-W., {Sohn}, S.~T., {et~al.} 2010, \aj, 139, 606

\bibitem[{{Cohen}(2004)}]{Cohen2004}
{Cohen}, J.~G. 2004, \aj, 127, 1545

\bibitem[{{D'Antona} {et~al.}(2016){D'Antona}, {Vesperini}, {D'Ercole},
  {Ventura}, {Milone}, {Marino}, \& {Tailo}}]{DAntona2016}
{D'Antona}, F., {Vesperini}, E., {D'Ercole}, A., {et~al.} 2016, \mnras, 458,
  2122

\bibitem[{{Dotter} {et~al.}(2008){Dotter}, {Chaboyer}, {Jevremovi{\'c}},
  {Kostov}, {Baron}, \& {Ferguson}}]{Dotter2008}
{Dotter}, A., {Chaboyer}, B., {Jevremovi{\'c}}, D., {et~al.} 2008, \apjs, 178,
  89

\bibitem[{{Einasto}(1979)}]{Einasto1979}
{Einasto}, J. 1979, in IAU Symposium, Vol.~84, The Large-Scale Characteristics
  of the Galaxy, ed. W.~B. {Burton}, 451--458

\bibitem[{{Eisenstein} {et~al.}(2011){Eisenstein}, {Weinberg}, {Agol},
  {Aihara}, {Allende Prieto}, {Anderson}, {Arns}, {Aubourg}, {Bailey},
  {Balbinot}, \& et~al.}]{Eisenstein2011}
{Eisenstein}, D.~J., {Weinberg}, D.~H., {Agol}, E., {et~al.} 2011, \aj, 142, 72

\bibitem[{{Fern{\'a}ndez-Alvar} {et~al.}(2018){Fern{\'a}ndez-Alvar}, {Carigi},
  {Schuster}, {Hayes}, {{\'A}vila-Vergara}, {Majewski}, {Allende Prieto},
  {Beers}, {S{\'a}nchez}, {Zamora}, {Garc{\'{\i}}a-Hern{\'a}ndez}, {Tang},
  {Fern{\'a}ndez-Trincado}, {Tissera}, {Geisler}, \& {Villanova}}]{FA2018}
{Fern{\'a}ndez-Alvar}, E., {Carigi}, L., {Schuster}, W.~J., {et~al.} 2018,
  \apj, 852, 50

\bibitem[{{Fern\'andez-Trincado}(2017)}]{Fernandez-Trincado2017Thesis}
{Fern\'andez-Trincado}, J.~G. 2017, Universit\'e Bourgogne Franche-Comt\'e Phd
  Thesis, 1, 174

\bibitem[{{Fern{\'a}ndez-Trincado}
  {et~al.}(2017{\natexlab{a}}){Fern{\'a}ndez-Trincado}, {Geisler}, {Moreno},
  {Zamora}, {Robin}, \& {Villanova}}]{FT2017c}
{Fern{\'a}ndez-Trincado}, J.~G., {Geisler}, D., {Moreno}, E., {et~al.}
  2017{\natexlab{a}}, ArXiv e-prints, arXiv:1710.07433

\bibitem[{{Fern{\'a}ndez-Trincado}
  {et~al.}(2017{\natexlab{b}}){Fern{\'a}ndez-Trincado}, {Robin}, {Moreno},
  {P{\'e}rez-Villegas}, \& {Pichardo}}]{Fernandez-Trincado2017b}
{Fern{\'a}ndez-Trincado}, J.~G., {Robin}, A.~C., {Moreno}, E.,
  {P{\'e}rez-Villegas}, A., \& {Pichardo}, B. 2017{\natexlab{b}}, ArXiv
  e-prints, arXiv:1708.05742

\bibitem[{{Fern{\'a}ndez-Trincado} {et~al.}(2016){Fern{\'a}ndez-Trincado},
  {Robin}, {Moreno}, {Schiavon}, {Garc{\'{\i}}a P{\'e}rez}, {Vieira}, {Cunha},
  {Zamora}, {Sneden}, {Souto}, {Carrera}, {Johnson}, {Shetrone}, {Zasowski},
  {Garc{\'{\i}}a-Hern{\'a}ndez}, {Majewski}, {Reyl{\'e}}, {Blanco-Cuaresma},
  {Martinez-Medina}, {P{\'e}rez-Villegas}, {Valenzuela}, {Pichardo}, {Meza},
  {M{\'e}sz{\'a}ros}, {Sobeck}, {Geisler}, {Anders}, {Schultheis}, {Tang},
  {Roman-Lopes}, {Mennickent}, {Pan}, {Nitschelm}, \&
  {Allard}}]{Fernandez-Trincado2016}
{Fern{\'a}ndez-Trincado}, J.~G., {Robin}, A.~C., {Moreno}, E., {et~al.} 2016,
  \apj, 833, 132

\bibitem[{{Fern{\'a}ndez-Trincado}
  {et~al.}(2017{\natexlab{c}}){Fern{\'a}ndez-Trincado}, {Zamora},
  {Garc{\'{\i}}a-Hern{\'a}ndez}, {Souto}, {Dell'Agli}, {Schiavon}, {Geisler},
  {Tang}, {Villanova}, {Hasselquist}, {Mennickent}, {Cunha}, {Shetrone},
  {Allende Prieto}, {Vieira}, {Zasowski}, {Sobeck}, {Hayes}, {Majewski},
  {Placco}, {Beers}, {Schleicher}, {Robin}, {M{\'e}sz{\'a}ros}, {Masseron},
  {Garc{\'{\i}}a P{\'e}rez}, {Anders}, {Meza}, {Alves-Brito}, {Carrera},
  {Minniti}, {Lane}, {Fern{\'a}ndez-Alvar}, {Moreno}, {Pichardo},
  {P{\'e}rez-Villegas}, {Schultheis}, {Roman-Lopes}, {Fuentes}, {Nitschelm},
  {Harding}, {Bizyaev}, {Pan}, {Oravetz}, {Simmons}, {Ivans},
  {Blanco-Cuaresma}, {Hern{\'a}ndez}, {Alonso-Garc{\'{\i}}a}, {Valenzuela}, \&
  {Chanam{\'e}}}]{Fernandez-Trincado2017a}
{Fern{\'a}ndez-Trincado}, J.~G., {Zamora}, O., {Garc{\'{\i}}a-Hern{\'a}ndez},
  D.~A., {et~al.} 2017{\natexlab{c}}, \apjl, 846, L2

\bibitem[{{Garc{\'{\i}}a-Hern{\'a}ndez}
  {et~al.}(2015){Garc{\'{\i}}a-Hern{\'a}ndez}, {M{\'e}sz{\'a}ros}, {Monelli},
  {Cassisi}, {Stetson}, {Zamora}, {Shetrone}, \& {Lucatello}}]{GH2015}
{Garc{\'{\i}}a-Hern{\'a}ndez}, D.~A., {M{\'e}sz{\'a}ros}, S., {Monelli}, M.,
  {et~al.} 2015, \apjl, 815, L4

\bibitem[{{Garc{\'{\i}}a P{\'e}rez} {et~al.}(2016){Garc{\'{\i}}a P{\'e}rez},
  {Allende Prieto}, {Holtzman}, {Shetrone}, {M{\'e}sz{\'a}ros}, {Bizyaev},
  {Carrera}, {Cunha}, {Garc{\'{\i}}a-Hern{\'a}ndez}, {Johnson}, {Majewski},
  {Nidever}, {Schiavon}, {Shane}, {Smith}, {Sobeck}, {Troup}, {Zamora},
  {Weinberg}, {Bovy}, {Eisenstein}, {Feuillet}, {Frinchaboy}, {Hayden},
  {Hearty}, {Nguyen}, {O\'{}Connell}, {Pinsonneault}, {Wilson}, \&
  {Zasowski}}]{GP2016}
{Garc{\'{\i}}a P{\'e}rez}, A.~E., {Allende Prieto}, C., {Holtzman}, J.~A.,
  {et~al.} 2016, \aj, 151, 144

\bibitem[{{Gonz{\'a}lez Hern{\'a}ndez} \& {Bonifacio}(2009)}]{GH2009}
{Gonz{\'a}lez Hern{\'a}ndez}, J.~I., \& {Bonifacio}, P. 2009, \aap, 497, 497

\bibitem[{{Gunn} {et~al.}(2006){Gunn}, {Siegmund}, {Mannery}, {Owen}, {Hull},
  {Leger}, {Carey}, {Knapp}, {York}, {Boroski}, {Kent}, {Lupton}, {Rockosi},
  {Evans}, {Waddell}, {Anderson}, {Annis}, {Barentine}, {Bartoszek}, {Bastian},
  {Bracker}, {Brewington}, {Briegel}, {Brinkmann}, {Brown}, {Carr},
  {Czarapata}, {Drennan}, {Dombeck}, {Federwitz}, {Gillespie}, {Gonzales},
  {Hansen}, {Harvanek}, {Hayes}, {Jordan}, {Kinney}, {Klaene}, {Kleinman},
  {Kron}, {Kresinski}, {Lee}, {Limmongkol}, {Lindenmeyer}, {Long}, {Loomis},
  {McGehee}, {Mantsch}, {Neilsen}, {Neswold}, {Newman}, {Nitta}, {Peoples},
  {Pier}, {Prieto}, {Prosapio}, {Rivetta}, {Schneider}, {Snedden}, \&
  {Wang}}]{Gunn2006}
{Gunn}, J.~E., {Siegmund}, W.~A., {Mannery}, E.~J., {et~al.} 2006, \aj, 131,
  2332

\bibitem[{{Harris}(1996)}]{Harris1996}
{Harris}, W.~E. 1996, \aj, 112, 1487

\bibitem[{{Hasselquist} {et~al.}(2017){Hasselquist}, {Shetrone}, {Smith},
  {Holtzman}, {McWilliam}, {Fern{\'a}ndez-Trincado}, {Beers}, {Majewski},
  {Nidever}, {Tang}, {Tissera}, {Fern{\'a}ndez Alvar}, {Allende Prieto},
  {Almeida}, {Anguiano}, {Battaglia}, {Carigi}, {Delgado Inglada},
  {Frinchaboy}, {Garc{\'{\i}}a-Hern{\'a}ndez}, {Geisler}, {Minniti}, {Placco},
  {Schultheis}, {Sobeck}, \& {Villanova}}]{Hasselquist2017}
{Hasselquist}, S., {Shetrone}, M., {Smith}, V., {et~al.} 2017, \apj, 845, 162

\bibitem[{{Hayes} {et~al.}(2018){Hayes}, {Majewski}, {Shetrone},
  {Fern{\'a}ndez-Alvar}, {Allende Prieto}, {Schuster}, {Carigi}, {Cunha},
  {Smith}, {Sobeck}, {Almeida}, {Beers}, {Carrera}, {Fern{\'a}ndez-Trincado},
  {Garc{\'{\i}}a-Hern{\'a}ndez}, {Geisler}, {Lane}, {Lucatello}, {Matthews},
  {Minniti}, {Nitschelm}, {Tang}, {Tissera}, \& {Zamora}}]{Hayes2018}
{Hayes}, C.~R., {Majewski}, S.~R., {Shetrone}, M., {et~al.} 2018, \apj, 852, 49

\bibitem[{{Holtzman} {et~al.}(2015){Holtzman}, {Shetrone}, {Johnson}, {Allende
  Prieto}, {Anders}, {Andrews}, {Beers}, {Bizyaev}, {Blanton}, {Bovy},
  {Carrera}, {Chojnowski}, {Cunha}, {Eisenstein}, {Feuillet}, {Frinchaboy},
  {Galbraith-Frew}, {Garc{\'{\i}}a P{\'e}rez}, {Garc{\'{\i}}a-Hern{\'a}ndez},
  {Hasselquist}, {Hayden}, {Hearty}, {Ivans}, {Majewski}, {Martell},
  {Meszaros}, {Muna}, {Nidever}, {Nguyen}, {O'Connell}, {Pan}, {Pinsonneault},
  {Robin}, {Schiavon}, {Shane}, {Sobeck}, {Smith}, {Troup}, {Weinberg},
  {Wilson}, {Wood-Vasey}, {Zamora}, \& {Zasowski}}]{Holtzman2015}
{Holtzman}, J.~A., {Shetrone}, M., {Johnson}, J.~A., {et~al.} 2015, \aj, 150,
  148

\bibitem[{{Howard} {et~al.}(2008){Howard}, {Rich}, {Reitzel}, {Koch}, {De
  Propris}, \& {Zhao}}]{Howard2008}
{Howard}, C.~D., {Rich}, R.~M., {Reitzel}, D.~B., {et~al.} 2008, \apj, 688,
  1060

\bibitem[{{Kharchenko} {et~al.}(2013){Kharchenko}, {Piskunov}, {Schilbach},
  {R{\"o}ser}, \& {Scholz}}]{Kharchenko2013}
{Kharchenko}, N.~V., {Piskunov}, A.~E., {Schilbach}, E., {R{\"o}ser}, S., \&
  {Scholz}, R.-D. 2013, \aap, 558, A53

\bibitem[{{Kunder} \& {Chaboyer}(2009)}]{Kunder2009}
{Kunder}, A., \& {Chaboyer}, B. 2009, \aj, 137, 4478

\bibitem[{{Kunder} {et~al.}(2012){Kunder}, {Koch}, {Rich}, {de Propris},
  {Howard}, {Stubbs}, {Johnson}, {Shen}, {Wang}, {Robin}, {Kormendy}, {Soto},
  {Frinchaboy}, {Reitzel}, {Zhao}, \& {Origlia}}]{Kunder2012}
{Kunder}, A., {Koch}, A., {Rich}, R.~M., {et~al.} 2012, \aj, 143, 57

\bibitem[{{Lauchner} {et~al.}(2006){Lauchner}, {Powell}, \&
  {Wilhelm}}]{Lauchner2006}
{Lauchner}, A., {Powell}, Jr., W.~L., \& {Wilhelm}, R. 2006, \apjl, 651, L33

\bibitem[{{Law} \& {Majewski}(2010)}]{Law2010}
{Law}, D.~R., \& {Majewski}, S.~R. 2010, \apj, 718, 1128

\bibitem[{{Libralato} {et~al.}(2018){Libralato}, {Bellini}, {Bedin}, {Moreno},
  {Fern{\'a}ndez-Trincado}, {Pichardo}, {van der Marel}, {Anderson}, {Apai},
  {Burgasser}, {Marino}, {Milone}, {Rees}, \& {Watkins}}]{Libralato2018}
{Libralato}, M., {Bellini}, A., {Bedin}, L.~R., {et~al.} 2018, ArXiv e-prints,
  arXiv:1801.01502

\bibitem[{{Majewski} {et~al.}(2017){Majewski}, {Schiavon}, {Frinchaboy},
  {Allende Prieto}, {Barkhouser}, {Bizyaev}, {Blank}, {Brunner}, {Burton},
  {Carrera}, {Chojnowski}, {Cunha}, {Epstein}, {Fitzgerald}, {Garc{\'{\i}}a
  P{\'e}rez}, {Hearty}, {Henderson}, {Holtzman}, {Johnson}, {Lam}, {Lawler},
  {Maseman}, {M{\'e}sz{\'a}ros}, {Nelson}, {Coung Nguyen}, {Nidever},
  {Pinsonneault}, {Shetrone}, {Smee}, {Smith}, {Stolberg}, {Skrutskie},
  {Walker}, {Wilson}, {Zasowski}, {Anders}, {Basu}, {Beland}, {Blanton},
  {Bovy}, {Brownstein}, {Carlberg}, {Chaplin}, {Chiappini}, {Eisenstein},
  {Elsworth}, {Feuillet}, {Fleming}, {Galbraith-Frew}, {Garc{\'{\i}}a},
  {An{\'{\i}}bal Garc{\'{\i}}a-Hern{\'a}ndez}, {Gillespie}, {Girardi}, {Gunn},
  {Hasselquist}, {Hayden}, {Hekker}, {Ivans}, {Kinemuchi}, {Klaene},
  {Mahadevan}, {Mathur}, {Mosser}, {Muna}, {Munn}, {Nichol}, {O'Connell},
  {Parejko}, {Robin}, {Rocha-Pinto}, {Schultheis}, {Serenelli}, {Shane}, {Silva
  Aguirre}, {Sobeck}, {Thompson}, {Troup}, {Weinberg}, \&
  {Zamora}}]{Majewski2017}
{Majewski}, S.~R., {Schiavon}, R.~P., {Frinchaboy}, P.~M., {et~al.} 2017, \aj,
  154, 94

\bibitem[{{Masseron} {et~al.}(2016){Masseron}, {Merle}, \&
  {Hawkins}}]{Masseron2016}
{Masseron}, T., {Merle}, T., \& {Hawkins}, K. 2016, {BACCHUS: Brussels
  Automatic Code for Characterizing High accUracy Spectra}, Astrophysics Source
  Code Library, , , ascl:1605.004

\bibitem[{{McLaughlin} \& {van der Marel}(2005)}]{McLaughlin2005}
{McLaughlin}, D.~E., \& {van der Marel}, R.~P. 2005, \apjs, 161, 304

\bibitem[{{M{\'e}sz{\'a}ros} {et~al.}(2015){M{\'e}sz{\'a}ros}, {Martell},
  {Shetrone}, {Lucatello}, {Troup}, {Bovy}, {Cunha},
  {Garc{\'{\i}}a-Hern{\'a}ndez}, {Overbeek}, {Allende Prieto}, {Beers},
  {Frinchaboy}, {Garc{\'{\i}}a P{\'e}rez}, {Hearty}, {Holtzman}, {Majewski},
  {Nidever}, {Schiavon}, {Schneider}, {Sobeck}, {Smith}, {Zamora}, \&
  {Zasowski}}]{Meszaros2015}
{M{\'e}sz{\'a}ros}, S., {Martell}, S.~L., {Shetrone}, M., {et~al.} 2015, \aj,
  149, 153

\bibitem[{{Milone} {et~al.}(2015){Milone}, {Marino}, {Piotto}, {Renzini},
  {Bedin}, {Anderson}, {Cassisi}, {D'Antona}, {Bellini}, {Jerjen},
  {Pietrinferni}, \& {Ventura}}]{Milone2015}
{Milone}, A.~P., {Marino}, A.~F., {Piotto}, G., {et~al.} 2015, \apj, 808, 51

\bibitem[{{Moreno} {et~al.}(2015){Moreno}, {Pichardo}, \&
  {Schuster}}]{Moreno2015}
{Moreno}, E., {Pichardo}, B., \& {Schuster}, W.~J. 2015, \mnras, 451, 705

\bibitem[{{Mottini} {et~al.}(2008){Mottini}, {Wallerstein}, \&
  {McWilliam}}]{Mottini2008}
{Mottini}, M., {Wallerstein}, G., \& {McWilliam}, A. 2008, \aj, 136, 614

\bibitem[{{Nidever} {et~al.}(2015){Nidever}, {Holtzman}, {Allende Prieto},
  {Beland}, {Bender}, {Bizyaev}, {Burton}, {Desphande}, {Fleming},
  {Garc{\'{\i}}a P{\'e}rez}, {Hearty}, {Majewski}, {M{\'e}sz{\'a}ros}, {Muna},
  {Nguyen}, {Schiavon}, {Shetrone}, {Skrutskie}, {Sobeck}, \&
  {Wilson}}]{Nidever2015}
{Nidever}, D.~L., {Holtzman}, J.~A., {Allende Prieto}, C., {et~al.} 2015, \aj,
  150, 173

\bibitem[{{Palma} {et~al.}(2002){Palma}, {Majewski}, \& {Johnston}}]{Palma2002}
{Palma}, C., {Majewski}, S.~R., \& {Johnston}, K.~V. 2002, \apj, 564, 736

\bibitem[{{Pancino} {et~al.}(2017){Pancino}, {Romano}, {Tang}, {Tautvai{\v
  s}ien{\.e}}, {Casey}, {Gruyters}, {Geisler}, {San Roman}, {Randich},
  {Alfaro}, {Bragaglia}, {Flaccomio}, {Korn}, {Recio-Blanco}, {Smiljanic},
  {Carraro}, {Bayo}, {Costado}, {Damiani}, {Jofr{\'e}}, {Lardo}, {de Laverny},
  {Monaco}, {Morbidelli}, {Sbordone}, {Sousa}, \& {Villanova}}]{Pancino2017}
{Pancino}, E., {Romano}, D., {Tang}, B., {et~al.} 2017, \aap, 601, A112

\bibitem[{{Pichardo} {et~al.}(2004){Pichardo}, {Martos}, \&
  {Moreno}}]{Pichardo2004}
{Pichardo}, B., {Martos}, M., \& {Moreno}, E. 2004, \apj, 609, 144

\bibitem[{{Piotto} {et~al.}(2015){Piotto}, {Milone}, {Bedin}, {Anderson},
  {King}, {Marino}, {Nardiello}, {Aparicio}, {Barbuy}, {Bellini}, {Brown},
  {Cassisi}, {Cool}, {Cunial}, {Dalessandro}, {D'Antona}, {Ferraro}, {Hidalgo},
  {Lanzoni}, {Monelli}, {Ortolani}, {Renzini}, {Salaris}, {Sarajedini}, {van
  der Marel}, {Vesperini}, \& {Zoccali}}]{Piotto2015}
{Piotto}, G., {Milone}, A.~P., {Bedin}, L.~R., {et~al.} 2015, \aj, 149, 91

\bibitem[{{Plez}(2012)}]{Plez2012}
{Plez}, B. 2012, {Turbospectrum: Code for spectral synthesis}, Astrophysics
  Source Code Library, , , ascl:1205.004

\bibitem[{{Portail} {et~al.}(2015){Portail}, {Wegg}, {Gerhard}, \&
  {Martinez-Valpuesta}}]{Portail2015}
{Portail}, M., {Wegg}, C., {Gerhard}, O., \& {Martinez-Valpuesta}, I. 2015,
  \mnras, 448, 713

\bibitem[{{Prantzos} {et~al.}(2007){Prantzos}, {Charbonnel}, \&
  {Iliadis}}]{Prantzos2007}
{Prantzos}, N., {Charbonnel}, C., \& {Iliadis}, C. 2007, \aap, 470, 179

\bibitem[{{Pryor} {et~al.}(2010){Pryor}, {Piatek}, \& {Olszewski}}]{Pryor2010}
{Pryor}, C., {Piatek}, S., \& {Olszewski}, E.~W. 2010, \aj, 139, 839

\bibitem[{{Recio-Blanco} {et~al.}(2017){Recio-Blanco}, {Rojas-Arriagada}, {de
  Laverny}, {Mikolaitis}, {Hill}, {Zoccali}, {Fern{\'a}ndez-Trincado}, {Robin},
  {Babusiaux}, {Gilmore}, {Randich}, {Alfaro}, {Allende Prieto}, {Bragaglia},
  {Carraro}, {Jofr{\'e}}, {Lardo}, {Monaco}, {Morbidelli}, \&
  {Zaggia}}]{Recio-Blanco2017}
{Recio-Blanco}, A., {Rojas-Arriagada}, A., {de Laverny}, P., {et~al.} 2017,
  \aap, 602, L14

\bibitem[{{Robin} {et~al.}(2017){Robin}, {Bienaym{\'e}},
  {Fern{\'a}ndez-Trincado}, \& {Reyl{\'e}}}]{Robin2017}
{Robin}, A.~C., {Bienaym{\'e}}, O., {Fern{\'a}ndez-Trincado}, J.~G., \&
  {Reyl{\'e}}, C. 2017, \aap, 605, A1

\bibitem[{{Robin} {et~al.}(2012){Robin}, {Marshall}, {Schultheis}, \&
  {Reyl{\'e}}}]{Robin2012}
{Robin}, A.~C., {Marshall}, D.~J., {Schultheis}, M., \& {Reyl{\'e}}, C. 2012,
  \aap, 538, A106

\bibitem[{{Robin} {et~al.}(2003){Robin}, {Reyl{\'e}}, {Derri{\`e}re}, \&
  {Picaud}}]{Robin2003}
{Robin}, A.~C., {Reyl{\'e}}, C., {Derri{\`e}re}, S., \& {Picaud}, S. 2003,
  \aap, 409, 523

\bibitem[{{Robin} {et~al.}(2014){Robin}, {Reyl{\'e}}, {Fliri}, {Czekaj},
  {Robert}, \& {Martins}}]{Robin2014}
{Robin}, A.~C., {Reyl{\'e}}, C., {Fliri}, J., {et~al.} 2014, \aap, 569, A13

\bibitem[{{Sarajedini} \& {Milone}(1995)}]{Sarajedini1995}
{Sarajedini}, A., \& {Milone}, A.~A.~E. 1995, \aj, 109, 269

\bibitem[{{Sbordone} {et~al.}(2007){Sbordone}, {Bonifacio}, {Buonanno},
  {Marconi}, {Monaco}, \& {Zaggia}}]{Sbordone2007}
{Sbordone}, L., {Bonifacio}, P., {Buonanno}, R., {et~al.} 2007, \aap, 465, 815

\bibitem[{{Sbordone} {et~al.}(2015){Sbordone}, {Monaco}, {Moni Bidin},
  {Bonifacio}, {Villanova}, {Bellazzini}, {Ibata}, {Chiba}, {Geisler},
  {Caffau}, \& {Duffau}}]{Sbordone2015}
{Sbordone}, L., {Monaco}, L., {Moni Bidin}, C., {et~al.} 2015, \aap, 579, A104

\bibitem[{{Schiavon} {et~al.}(2017){Schiavon}, {Johnson}, {Frinchaboy},
  {Zasowski}, {M{\'e}sz{\'a}ros}, {Garc{\'{\i}}a-Hern{\'a}ndez}, {Cohen},
  {Tang}, {Villanova}, {Geisler}, {Beers}, {Fern{\'a}ndez-Trincado},
  {Garc{\'{\i}}a P{\'e}rez}, {Lucatello}, {Majewski}, {Martell}, {O'Connell},
  {Prieto}, {Bizyaev}, {Carrera}, {Lane}, {Malanushenko}, {Malanushenko},
  {Mu{\~n}oz}, {Nitschelm}, {Oravetz}, {Pan}, {Roman-Lopes}, {Schultheis}, \&
  {Simmons}}]{Schiavon2017}
{Schiavon}, R.~P., {Johnson}, J.~A., {Frinchaboy}, P.~M., {et~al.} 2017,
  \mnras, 466, 1010

\bibitem[{{Schlafly} \& {Finkbeiner}(2011)}]{Schlafly2011}
{Schlafly}, E.~F., \& {Finkbeiner}, D.~P. 2011, \apj, 737, 103

\bibitem[{{Shetrone} {et~al.}(2015){Shetrone}, {Bizyaev}, {Lawler}, {Allende
  Prieto}, {Johnson}, {Smith}, {Cunha}, {Holtzman}, {Garc{\'{\i}}a P{\'e}rez},
  {M{\'e}sz{\'a}ros}, {Sobeck}, {Zamora}, {Garc{\'{\i}}a-Hern{\'a}ndez},
  {Souto}, {Chojnowski}, {Koesterke}, {Majewski}, \& {Zasowski}}]{Shetrone2015}
{Shetrone}, M., {Bizyaev}, D., {Lawler}, J.~E., {et~al.} 2015, \apjs, 221, 24

\bibitem[{{Smith} {et~al.}(2015){Smith}, {Flynn}, {Candlish}, {Fellhauer}, \&
  {Gibson}}]{Smith2015}
{Smith}, R., {Flynn}, C., {Candlish}, G.~N., {Fellhauer}, M., \& {Gibson},
  B.~K. 2015, \mnras, 448, 2934

\bibitem[{{Sofue}(2015)}]{Sofue2015}
{Sofue}, Y. 2015, \pasj, 67, 75

\bibitem[{{Sohn} {et~al.}(2012){Sohn}, {Anderson}, \& {van der
  Marel}}]{Sohn2012}
{Sohn}, S.~T., {Anderson}, J., \& {van der Marel}, R.~P. 2012, \apj, 753, 7

\bibitem[{{Sohn} {et~al.}(2013){Sohn}, {Besla}, {van der Marel},
  {Boylan-Kolchin}, {Majewski}, \& {Bullock}}]{Sohn2013}
{Sohn}, S.~T., {Besla}, G., {van der Marel}, R.~P., {et~al.} 2013, \apj, 768,
  139

\bibitem[{{Sohn} {et~al.}(2017){Sohn}, {Patel}, {Besla}, {van der Marel},
  {Bullock}, {Strigari}, {van de Ven}, {Walker}, \& {Bellini}}]{Sohn2017}
{Sohn}, S.~T., {Patel}, E., {Besla}, G., {et~al.} 2017, \apj, 849, 93

\bibitem[{{Souto} {et~al.}(2016){Souto}, {Cunha}, {Smith}, {Allende Prieto},
  {Pinsonneault}, {Zamora}, {Garc{\'{\i}}a-Hern{\'a}ndez}, {M{\'e}sz{\'a}ros},
  {Bovy}, {Garc{\'{\i}}a P{\'e}rez}, {Anders}, {Bizyaev}, {Carrera},
  {Frinchaboy}, {Holtzman}, {Ivans}, {Majewski}, {Shetrone}, {Sobeck}, {Pan},
  {Tang}, {Villanova}, \& {Geisler}}]{Souto2016}
{Souto}, D., {Cunha}, K., {Smith}, V., {et~al.} 2016, \apj, 830, 35

\bibitem[{{Tang} {et~al.}(2017){Tang}, {Cohen}, {Geisler}, {Schiavon},
  {Majewski}, {Villanova}, {Carrera}, {Zamora}, {Garcia-Hernandez}, {Shetrone},
  {Frinchaboy}, {Meza}, {Fern{\'a}ndez-Trincado}, {Mu{\~n}oz}, {Lin}, {Lane},
  {Nitschelm}, {Pan}, {Bizyaev}, {Oravetz}, \& {Simmons}}]{Tang2017}
{Tang}, B., {Cohen}, R.~E., {Geisler}, D., {et~al.} 2017, \mnras, 465, 19

\bibitem[{{Ventura} {et~al.}(2012){Ventura}, {D'Antona}, {Di Criscienzo},
  {Carini}, {D'Ercole}, \& {vesperini}}]{Ventura2012}
{Ventura}, P., {D'Antona}, F., {Di Criscienzo}, M., {et~al.} 2012, \apjl, 761,
  L30

\bibitem[{{Zamora} {et~al.}(2015){Zamora}, {Garc{\'{\i}}a-Hern{\'a}ndez},
  {Allende Prieto}, {Carrera}, {Koesterke}, {Edvardsson}, {Castelli}, {Plez},
  {Bizyaev}, {Cunha}, {Garc{\'{\i}}a P{\'e}rez}, {Gustafsson}, {Holtzman},
  {Lawler}, {Majewski}, {Manchado}, {M{\'e}sz{\'a}ros}, {Shane}, {Shetrone},
  {Smith}, \& {Zasowski}}]{Zamora2015}
{Zamora}, O., {Garc{\'{\i}}a-Hern{\'a}ndez}, D.~A., {Allende Prieto}, C.,
  {et~al.} 2015, \aj, 149, 181

\bibitem[{{Zasowski} {et~al.}(2017){Zasowski}, {Cohen}, {Chojnowski},
  {Santana}, {Oelkers}, {Andrews}, {Beaton}, {Bender}, {Bird}, {Bovy},
  {Carlberg}, {Covey}, {Cunha}, {Dell'Agli}, {Fleming}, {Frinchaboy},
  {Garcia-Hernandez}, {Harding}, {Holtzman}, {Johnson}, {Kollmeier},
  {Majewski}, {Meszaros}, {Munn}, {Munoz}, {Ness}, {Nidever}, {Poleski},
  {Zuniga}, {Shetrone}, {Simon}, {Smith}, {Sobeck}, {Stringfellow},
  {Szigetiaros}, {Tayar}, \& {Troup}}]{Zasowski2017}
{Zasowski}, G., {Cohen}, R.~E., {Chojnowski}, S.~D., {et~al.} 2017, ArXiv
  e-prints, arXiv:1708.00155

\end{thebibliography}

\label{lastpage}

\end{document}